\documentclass[final,1p,times]{elsarticle}

\usepackage{graphicx}
\usepackage{float} 

\usepackage[intlimits]{amsmath}
\usepackage{amsfonts,amssymb}

\biboptions{numbers, sort&compress}

\usepackage{hyperref}
\hypersetup{colorlinks=true, citecolor=blue}
\usepackage[hypcap=true]{caption}

\usepackage{color}

\usepackage{upgreek}

\newcommand{\figurepanel}[2]{\hyperref[#1]{\ref*{#1}(#2)}}

\allowdisplaybreaks[2] 

\usepackage[T1]{fontenc} 

\makeatletter

\usepackage{cancel}

\usepackage{rotating} 
\usepackage{tablefootnote}

\DeclareMathOperator{\re}{Re}
\DeclareMathOperator{\im}{Im}

\usepackage{filecontents}

\newcounter{bla}

\journal{Computer Physics Communications}

\begin{document}

\begin{frontmatter}



\title{FDTD: solving 1+1D delay PDE in parallel}


\author[e,g,h]{Yao-Lung L. Fang}

\address[e]{Department of Physics, Duke University, P.O.\ Box 90305, Durham, North Carolina 27708-0305, USA}
\address[g]{Computational Science Initiative, Brookhaven National Laboratory, Upton, NY 11973-5000, USA\footnotemark[1]}
\address[h]{National Synchrotron Light Source II, Brookhaven National Laboratory, Upton, NY 11973-5000, USA\footnotemark[1]}
\footnotetext[1]{Present address.}

\begin{abstract}
We present a proof of concept for solving a 1+1D complex-valued, delay partial differential equation (PDE) that emerges in the study of waveguide quantum electrodynamics (QED) by adapting the finite-difference time-domain (FDTD) method. The delay term is spatially non-local, rendering conventional approaches such as the method of lines inapplicable. 
We show that by properly designing the grid and by supplying the (partial) exact solution as the boundary condition, the delay PDE can be numerically solved. In addition, we demonstrate that while the delay imposes strong data dependency, multi-thread parallelization can nevertheless be applied to such a problem. Our code provides a numerically exact solution to the time-dependent multi-photon scattering problem in waveguide QED. 
\end{abstract}

\begin{keyword}
Waveguide QED \sep Delay PDE \sep FDTD \sep Non-Markovianity
\end{keyword}

\end{frontmatter}



{\bf PROGRAM SUMMARY}

\begin{small}
\noindent
{\em Program Title:} FDTD: solving 1+1D delay PDE                                          \\
{\em Licensing provisions:} MIT                                  \\
{\em Programming language:} C (C99)                                  \\

{\em Supplementary material:}                               \\
{\em Journal reference of previous version:}                  \\
{\em Does the new version supersede the previous version?:}   \\
{\em Reasons for the new version:}\\
{\em Summary of revisions:}*\\

{\em Nature of problem(approx. 50-250 words):}
This program solves an unconventional 1+1D delay PDE that emerges in the study of waveguide quantum electrodynamics. The delay PDE is complex-valued and has a non-local delay term, and the solution to it provides the full dynamics of the system consisting of a few 1D photons and a two-level system in front of a mirror.\\

{\em Solution method(approx. 50-250 words):}
The finite-difference time-domain (FDTD) method is adapted. Given the initial condition of the system, the corresponding boundary condition is generated, and then the FDTD solver marches through the entire spacetime grid. Multiple solvers are supported using either OpenMP (wavefront) or pthreads (swarm).\\

{\em Additional comments including Restrictions and Unusual features (approx. 50-250 words):}
1.\ Depending on the input parameters the memory and disk usages of the program can be excessive, so the users should choose the parameters wisely (see main text).
2.\ The multi-thread support using OpenMP is turned on by default. See README for how to turn it off and switch to pthreads instead.
3.\ As a by-product, a numerical routine is provided for evaluating the incomplete Gamma function $\gamma(n, z)$ with nonzero positive integers $n\geq1$ and complex-valued $z$.
\\
   \\

\end{small}

\section{Introduction}
Waveguide quantum electrodynamics (QED) concerns the interaction between one-dimensional (1D) waveguide photons and local emitters (atoms, qubits, etc.) \cite{LodahlRMP15,RoyRMP17,NohRPP16,LiaoPhyScr16,GuPR17}. In the cases where a coherent feedback loop is formed due to the presence of, for example, multiple distant emitters or a perfect mirror terminating the waveguide, the propagation of photons needs to be taken into account rigorously if the time of flight between distant objects is non-negligible compared to the decay time of the qubits \cite{ZhengPRL13,TufarelliPRA14,FangPRA15,GrimsmoPRL15,RamosPRA16,PichlerPRL16}.

The study of time evolution (i.e., dynamics) in waveguide QED has drawn considerable attention \cite{LongoPRL10, LongoPRA11, PeropadrePRL13, 
	ZuecoFD14,ShiPRA15, KocabasPRA16, KocabasOL16, MirzaPRA16b, GuoPRA17, WhalenQST17,GuimondQST17}, 
since it offers full information of light-matter interaction based on which more precise control of the system or detailed analysis of quantum non-Markovianity \cite{RivasRPP14,BreuerRMP16,deVegaRMP17} can be performed. Insight into the non-equilibrium physics can also be obtained. While powerful numerical approaches, such as density matrix renormalization group (DMRG) \cite{WhitePRL92,SchollwockRMP05,SchollwockAnnPhys11} whose accuracy for solving 1D systems is undoubtedly demonstrated, have been applied recently to waveguide QED problems with feedback loops \cite{PichlerPRL16,GuimondQST17}, in the present paper and Ref.~\cite{FangNJP18} we show that simply solving the \emph{time-dependent} Schr\"{o}dinger equation can provide a perhaps more natural and intuitive viewpoint on the scattering physics.
However, in the presence of delay, the Schr\"{o}dinger equation leads to a delay partial differential equation (PDE), solving which is a significant technical challenge to be addressed by this paper.

As a concrete example, we consider a two-level system (2LS) coupled to a semi-infinite waveguide, one end of which is terminated by a perfect mirror \cite{FangPRA15, FangNJP18}. 
Under the rotating-wave approximation, the number of excitations in a waveguide-QED system is conserved. Therefore, one may partition the full Hilbert space into different number sectors. While the dynamics in the one-excitation sector is described by a delay differential equation (DDE)\footnote{To be more consistent with respect to its PDE counterpart, one could use the acronym ``delay ODE'', which stands for \emph{delay ordinary differential equation}. However, we use DDE to follow the convention in the literature.}, a one-variable ordinary differential equation that has a delay term \cite{FangNJP18, TufarelliPRA13, DornerPRA02} and can be solved straightforwardly \cite{DDEbook}, the multi-excitation sectors require nontrivial care. In particular, in the \emph{two-excitation} sector there can be either two propagating photons in the waveguide (wavefunction denoted by $\chi$ hereafter), or one flying and the other absorbed by the 2LS ($\psi$); note that a 2LS can only hold one photon at a time.
Therefore, starting from the Schr\"{o}dinger equation and unfolding the half space, we arrive at
a 1+1D delay PDE that describes the (complex-valued) time-dependent wavefunction, $\psi(x,t)$, of the 2LS plus a photon at position $x$:
\begin{equation}
\begin{aligned}
\frac{\partial}{\partial t}\psi(x,t)&=-\frac{\partial}{\partial x}\psi(x,t)-\left(i\omega_0+\frac{\Gamma}{2}\right)\psi(x,t)+\frac{\Gamma}{2}\psi(x-2a, t-2a)\theta(t-2a)  \\
&\quad- \frac{\Gamma}{2}\Biggl\{\Bigl[\psi(- x- 2a, t- x- a) - \psi(- x, t- x- a)\Bigr]\theta(x+a)\theta(t- x- a)\\
&\quad\quad\quad +\Bigl[\psi(2a- x, t- x+ a)- \psi(- x, t- x+a)\Bigr]\theta(x- a)\theta(t- x+ a)\Biggr\} \\
&\quad+\sqrt{\Gamma}\biggl[\chi(x-t,-a-t,0)-\chi(x-t,a-t,0)\biggr],
\end{aligned}\label{eq:double-excitation delay differential eq}
\end{equation}
where $a$ is the atom-mirror separation, $\omega_0$ ($\Gamma$) is the 2LS frequency (decay rate), $\theta(x)$ is the step function, $\chi(x_1, x_2,t )$ is the time-dependent two-photon wavefunction (one at position $x_1$, another at $x_2$), and we set $\hbar=c=1$. By solving Eq.~\eqref{eq:double-excitation delay differential eq} for $\psi$, the full dynamics of the system can be completely determined, since the
two-photon wavefunction 
can be written in terms of $\psi$:  
\begin{equation}
\begin{aligned}
\chi(x_1, x_2, t)& = \chi(x_1-t, x_2-t, 0)-\frac{\sqrt{\Gamma}}{2}\biggl[
\psi(x_1-x_2-a, t-x_2-a)\theta(x_2+a)\theta(t-x_2-a)\\
&\quad
-\psi(x_1-x_2+a, t-x_2+a)\theta(x_2-a)\theta(t-x_2+a) +\bigl(x_2 \leftrightarrow x_1\bigr)
\biggr].
\label{eq:formal solution of two-photon wavefunction}
\end{aligned}
\end{equation}
Note that $\chi$ is symmetric under the exchange of $x_1$ and $x_2$ due to bosonic statistics. The detailed discussion of this problem and the derivation of above equations are reported in \cite{FangNJP18}.

Before proceeding, we first reiterate that the focus of the present paper is on delay PDE, not DDE. As stated above, solving DDE subject to appropriate initial conditions, in particular linear DDE of the form
$\dot{x}(t) = A x(t-\uptau) + B x(t) + C$,
is simple and in fact standard \cite{DDEbook}. 
The solution usually consists of an infinite sum of piecewise functions whose interval is dictated by the delay $\uptau$. For solving more complicated cases, such as a nonlinear DDE, DDE with time-dependent coefficients, a system of DDEs, and possibly a combination of these, there are also sophisticated numerical routines provided in, e.g., Matlab and Mathematica.
Second, while delay PDEs have emerged in some scientific and engineering contexts and been numerically studied using, for example, the method of lines \cite{Zubik_Kowal_Sch08}, we emphasize that as far as we understand, those approaches cannot be directly applied to our problem, as the delay term in Eq.~\eqref{eq:double-excitation delay differential eq} lies in both $x$ and $t$ dimensions (i.e., it's a non-local delay), contrary to the common situation of delay PDE in which only one of the dimensions is delayed \cite{Zubik_Kowal_Sch08}. In other words, Eq.~\eqref{eq:double-excitation delay differential eq} cannot be converted to a system of ordinary differential equations (ODE) that is discretized in $x$, and then be solved by an ODE/DDE solver along $t$. Furthermore, to the best of our knowledge there is no general-purpose solvers for delay PDE. As a result, we adapt and implement the finite-difference time-domain (FDTD) method for solving the delay PDE \eqref{eq:double-excitation delay differential eq} and demonstrate its validity in this paper.  

Moreover, because of the feedback loop the system is highly ``non-Markovian'', a jargon widely used in the community of open quantum systems \cite{RivasRPP14,BreuerRMP16,deVegaRMP17} meaning the quantum system has a dependence on its past history. Since solving the wavefunction requires looking up the system's memory (values of $\psi$ solved at earlier times), as we will see this imposes stringent constraints on how the problem can be parallelized. In the paper, we present, implement, and benchmark two different multi-thread approaches, \emph{swarm} and \emph{wavefront}, to address this issue. 

The main purpose of this work is therefore three-fold: (i) to numerically solve Eq.~\eqref{eq:double-excitation delay differential eq} using FDTD; (ii) to provide a proof of concept that FDTD works well for tackling complex-valued, spatially non-local delay PDE and that certain degree of parallelism can be achieved; (iii)  to present a numerically exact solution to the \emph{time-dependent, multi-photon scattering} problem in waveguide QED.

\section{Method and targeted problems}
FDTD is widely used by engineers in antenna designs, computational electrodynamics, plasmonics, etc. 
Below we briefly discuss how FDTD works, and refer interested readers to Refs.~\cite{FDTDbook,FDTDnote} for the details.

Like most of differential-equation solving methods, FDTD discretizes the spacetime, and different discretization schemes have their own advantages and disadvantages. For simplicity we choose a square lattice, and note in passing that in real FDTD applications the Yee (staggered) lattice is more common, as the conservation laws of EM fields are trivially hold on the Yee lattice. In the following we set the ratio of the spatial step $\Delta_x$  to the temporal step $\Delta_t$  equal to the speed of light $c=1$ so that $\Delta_t=\Delta_x=\Delta$. In FDTD, $\Delta_x/\Delta_t\geq c$ is called the Courant condition and is hold when the algorithm is stable. See \ref{appen: stability} for a simplified discussion on stability.

We next express every term in Eq.~\eqref{eq:double-excitation delay differential eq} using finite differences. We use the ``leapfrog'' prescription that has $\mathcal{O}(\Delta^2)$ accuracy \cite{NumericalRecipes}:
\begin{subequations}
\begin{align}
\frac{\partial f(x+\Delta/2,t+\Delta/2)}{\partial t} & \approx\frac{1}{\Delta} \biggl(\frac{f(x,t+\Delta)+f(x+\Delta,t+\Delta)}{2}
-\frac{f(x,t)+f(x+\Delta,t)}{2}\biggr)\\
\frac{\partial f(x+\Delta/2,t+\Delta/2)}{\partial x} &\approx \frac{1}{\Delta} \biggl(\frac{f(x+\Delta,t+\Delta)+f(x+\Delta,t)}{2}
-\frac{f(x,t)+f(x,t+\Delta)}{2} \biggr)\\
f(x+\Delta/2,t+\Delta/2)&\approx \frac{1}{4}\biggl(f(x+\Delta,t+\Delta)+f(x+\Delta,t)
+f(x,t+\Delta)+f(x,t)\biggr)
\end{align}
	\label{eq:square average}
\end{subequations}
In the code we call Eq.~\eqref{eq:square average} a ``square'', because the value at the center of a square is approximated using the values at its four corners. These discretization rules apply to the terms in the first line of Eq.~\eqref{eq:double-excitation delay differential eq} so that given the values at three corners, the value at the top right corner of a square can be solved. As for those terms in the second and the third lines of Eq.~\eqref{eq:double-excitation delay differential eq}, hereafter referred to as the source terms \cite{FangNJP18}, we need a different representation:
\begin{equation}
f(x+\Delta/2,t)\approx \frac{1}{2}\biggl( f(x,t) + f(x+\Delta,t)\biggr), \label{eq:bar average}
\end{equation}
which we call a ``bar''. The reason for using bars over squares will become clear shortly.


\begin{figure}[h]
	\centering
	\includegraphics[scale=0.6]{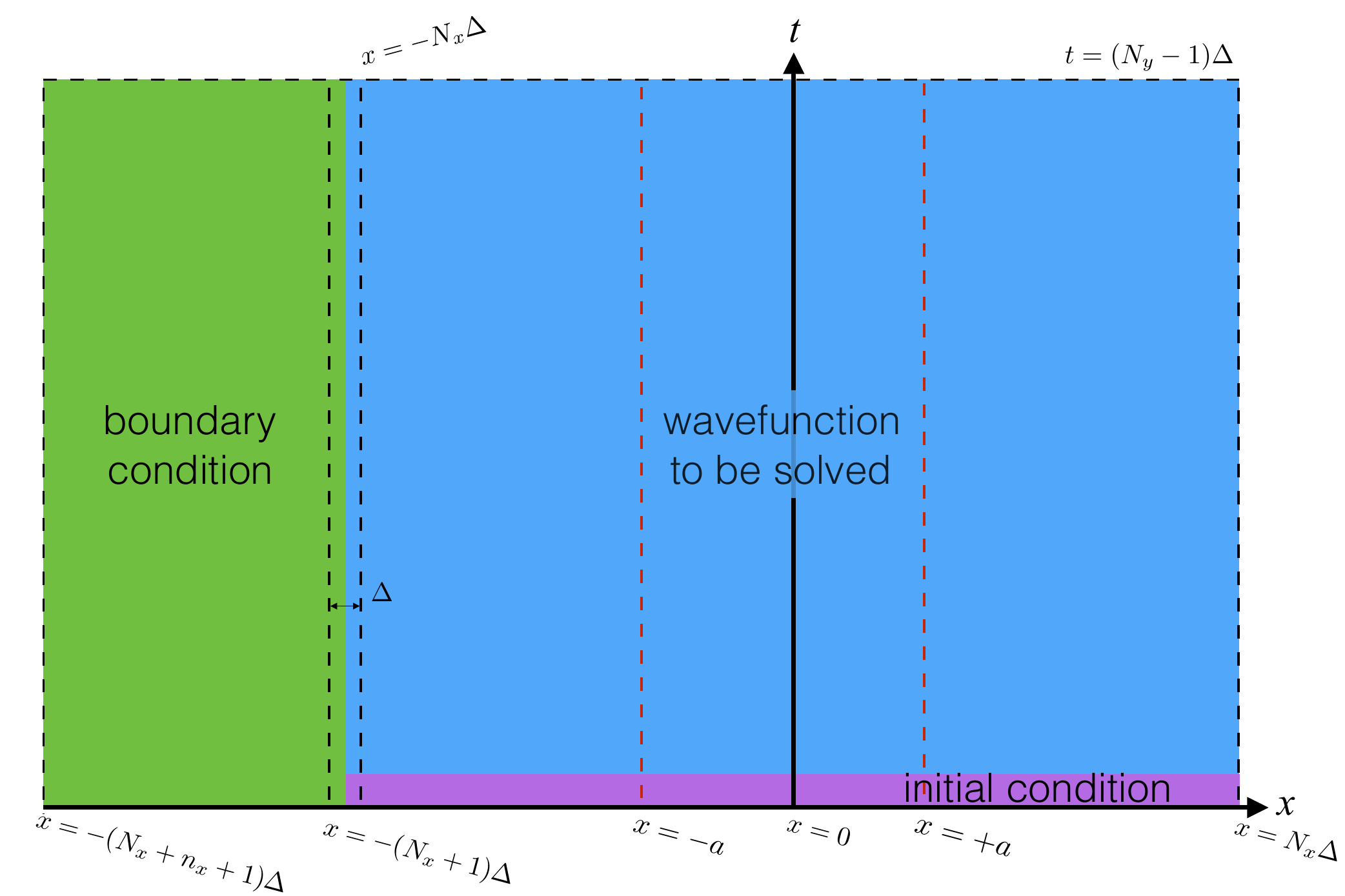}
	\caption{The schematic of the spacetime layout used in FDTD. Note that the lines $x=\pm a=\pm (n_x/2)\Delta$ locate in the blue region if $n_x\leq2N_x$, and that $x=-a$ is at the center of the grid.}
	\label{fig:FDTD_schematic}
\end{figure}

\begin{sidewaystable} 
	\centering
	\caption{\label{table:input} Summary of all input parameters accepted by our program.}
	\begin{tabular}{|c| p{8cm} |c| c|}
		\hline
		\textbf{Parameter} &  \textbf{Description} & \textbf{Mandatory} & \textbf{Default} \\ \hline
		\texttt{Nx} & Defined such that total grid points (to be solved) along $x$ is $2N_x+1$ (so $x/\Delta\in[-N_x,N_x]$) & Yes & n/a \\ \hline
		\texttt{Ny} & Total grid points along $t$ (so $t/\Delta\in[0,N_y-1]$) & Yes & n/a \\ \hline
		\texttt{nx} & $n_x = 2a/\Delta$; $n_x$ needs to be an integer multiple of 2 and $n_x\leq2N_x$ & Yes & n/a \\ \hline
		\texttt{Delta} & Step size $\Delta$ (See also \ref{appen: translation}) & Yes &  n/a\\ \hline
		\texttt{k} & Driving frequency $k$ (in units of $\Delta^{-1}$) & Yes & 0 \\ \hline
		\texttt{k1}, \texttt{k2} & Incident frequencies for photon \#1 (\#2) (in units of $\Delta^{-1}$) & No\tablefootnote{Needed when \texttt{init\_cond=3} and \texttt{identical\_photons=0}.\label{note6}}  & 0 \\ \hline
		\texttt{w0} & 2LS frequency $\omega_0$ (in units of $\Delta^{-1}$) & Yes & n/a\\ \hline
		\texttt{gamma} & 2LS decay rate $\Gamma$ (in units of $\Delta^{-1}$) & Yes & n/a\\ \hline
		\texttt{init\_cond} & 1: two-photon plane wave; 2: stimulated emission (one-photon exponential wavepacket); 3: two-photon exponential wavepacket & Yes & 0 \\ \hline
		\texttt{alpha} & Wavepacket width $\alpha$ (in units of $\Gamma$) & No\tablefootnote{Needed when \texttt{init\_cond=2}, and ineffective when \texttt{init\_cond=1}.} & 0 \\ \hline
		\texttt{alpha1}, \texttt{alpha2} & Wavepacket width for photon \#1 (\#2) (in units of $\Gamma$) & No\textsuperscript{\ref{note6}}& 0 \\ \hline
		\texttt{identical\_photons} & whether or not the two incident photons are the same & No & 1 \\ \hline
		\texttt{save\_chi} & Output $\chi(a+\Delta, a+\Delta+\uptau, t)$ as plain text & No\tablefootnote{These options cannot be simultaneously turned off (set to 0), or no output will be generated.\label{note2}} & 0 \\ \hline
		\texttt{save\_psi} & Output $\psi(x, t)$ as plain text & No\textsuperscript{\ref{note2}} & 0 \\ \hline
		\texttt{save\_psi\_binary} & Output $\psi(x, t)$ as binary & No\textsuperscript{\ref{note2}} & 0 \\ \hline
		\texttt{save\_psi\_square\_integral} & Output $\int dx\,|\psi(x, t)|^2$ as plain text & No\textsuperscript{\ref{note2}} & 0 \\ \hline
		\texttt{measure\_NM} & See \ref{append: NM} & No\textsuperscript{\ref{note2},}\tablefootnote{Requires \texttt{init\_cond=2}.} & 0\\ \hline
		\texttt{Tstep} & Output the wavefunctions for every $\texttt{Tstep}+1$ temporal steps& No & 0 \\ \hline
		\texttt{Nth} &  Number of solvers & No & 1 \\ \hline
	\end{tabular}
\end{sidewaystable}

Now we can put together squares and bars to discretize Eq.~\eqref{eq:double-excitation delay differential eq}. For putting the problem on a computer, we also need to draw a ``box'' as we cannot  walk through the entire spacetime indefinitely. As a result, we need to specify 4 parameters to define the box geometry: 
$N_x$, $N_y$, $n_x$, and $\Delta$;
see Table~\ref{table:input} for the list of accepted input parameters.
The corresponding layout is shown in Fig.~\ref{fig:FDTD_schematic}. Note that (a) the initial condition is given on the line $t=0$ (the purple stripe); (b) the boundary condition is given for not just one line, as needed for solving ordinary (space-local and time-local) PDEs, but for a wide area (the green region) because of the delay and source terms; (c) in order to reach $x\geq+a$ and to make the layout well-defined, we need $n_x$ to be an integer multiple of 2 and $n_x\leq2N_x$; (d) the step size $\Delta$ can be given \emph{arbitrarily}, see \ref{appen: translation}.

\begin{figure}[htb]
	\centering
	\includegraphics[scale=0.35]{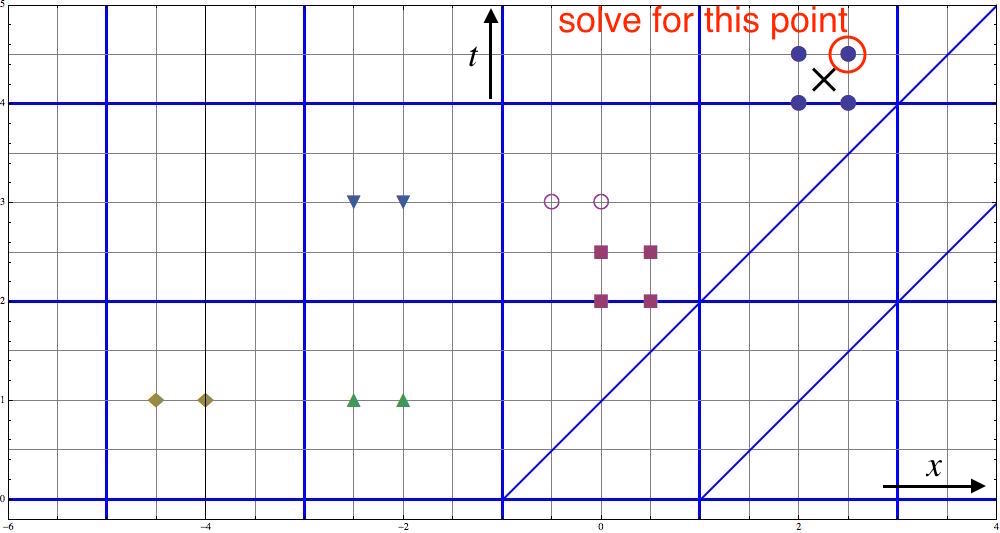}
	\caption{An example of calculating $\psi(x,t)$ for the circled point on the square lattice. The slant lines represent light cones extended from the coupling points at $x=\pm a$. $n_x$ is chosen to be 4 for illustrative purposes, and the black cross denotes the Taylor-expansion point, referred to as the solver. The other colored points contribute to the point to be solved [each color corresponds to a term in Eq.~\eqref{eq:double-excitation delay differential eq}].}
	\label{fig:FDTD_grid}
\end{figure}

It is illustrative to see how the FDTD solves Eq.~\eqref{eq:double-excitation delay differential eq}. Fig.~\ref{fig:FDTD_grid} shows a snapshot of the FDTD solver which marches in a space-then-time manner. Since the delay PDE is \emph{chiral} (unidirectional) \cite{FangNJP18}, for a given time $t$ the solver (conceptually represented by the black cross) moves from the left edge of the box to the right, then advances one step $\Delta$ in time and repeats. Each term in Eq.~\eqref{eq:double-excitation delay differential eq} has a different color for easy identification, and we use all previously solved values to solve for the top-right corner of the blue square, which is circled in red. Note that there are four colors, each of which has only two points (the bars), because when we Taylor-expand at the black cross, those terms are expanded at the center between the two points. 

Furthermore, from Fig.~\ref{fig:FDTD_grid} one can appreciate the fact that in general the system is highly  non-Markovian.
In terms of programming, this brings in a heavy burden because, for performance reasons, one can no longer flush earlier values from memory to disk, as typically done in solving ordinary PDEs. Instead, one needs to keep all grid points in the memory, so the hardware capacity is an important factor; doing frequent I/O is not an efficient option. 
For example, as we are solving a complex wavefunction,
depending on the grid size it can be very memory-intensive (each grid point stores a complex double number and thus takes 16 bytes)
and space-intensive (the wavefunction is written to a plain txt when the calculation is done). Readers should use the program with caution.\footnote{A quick estimation for memory usage is roughly $32N_xN_y/1024^3$ (in GB), and for disk usage divided by $\texttt{Tstep}+1$.}

Currently the FDTD program can solve three classes of problems, and for each class the boundary condition (green in Fig.~\ref{fig:FDTD_schematic}) is given by known analytical expressions:
\begin{enumerate}
	\item[(a)] \textit{Two-photon plane wave} \cite{FangNJP18}: the incoming photons are described by continuous wave, $\chi(x_1, x_2, 0) = A^2 \exp[i k (x_1+x_2)]\theta(-a-x_1)\theta(-a-x_2)$, and the initial condition for $\psi$ is simply $\psi(x,0)=0$ (the 2LS is initially in its ground state). Therefore, we need to supply three more parameters that are physics-related: $k$, $\omega_0$, and $\Gamma$; see Table~\ref{table:input}. 
	In $x<-a$, the solution to Eq.~\eqref{eq:double-excitation delay differential eq} is 
	\begin{equation}
	\psi(x,t) = \sqrt{2}Ae^{ik(x-t)}e_0(t), 
	\label{eq: BC two-photon scattering}
	\end{equation}
	where $e_0(t)$ is the 2LS wavefunction, solved in the one-excitation sector assuming $e_0(0)=0$ and $\phi(x, 0)=A e^{ikx}\theta(-a-x)$,
	\begin{equation}
	\begin{aligned}
	e_0(t)&=\frac{i\sqrt{\frac{\Gamma}{2}} Ae^{-ika}(e^{-ikt}-e^{-(i\omega_0+\Gamma/2)t})}{p}
    - Ae^{-ika} \sum_{n=1}^{\infty}\frac{\left(\frac{\Gamma}{2}\right)^{n-1/2}}{n!}
	\Bigl[(t-2na)^n e^{-(i\omega_0+\Gamma/2)(t-2na)}\\
	&\quad\quad+\frac{i^n(k-\omega_0)}{p^{n+1}}\gamma(n+1,-ip(t-2na))e^{-ik(t-2na)}
	\Bigr]\theta(t-2na),
	\end{aligned}\label{eq:exact solution single-photon scattering}
	\end{equation}
	where $p=k-\omega_0+i\Gamma/2$, and $\gamma(n, z)$ is the (lower) incomplete Gamma function \cite{NISThandbook}. We note that evaluating $\gamma(n, z)$ on the complex plane is in general a non-trivial task; see \ref{appen: incom_gamma}.
	The above expressions are used to generate both the initial and boundary conditions. Finally, we note that $A=1$ is set in the program for convenience.
	
	\item[(b)] \textit{Stimulated emission} \cite{FangNJP18}: a single-photon exponential wavepacket of the form\footnote{The initial wavefront position is set at $x=-a$ to shorten the computation time and to maximize interference effects, otherwise we need to wait for the wavepacket to arrive, and by then the qubit may already decay.}
	\begin{equation}
	\varphi(x)=i\sqrt{\alpha \Gamma} e^{ikx+\alpha\Gamma(x+a)/2}\theta(-x-a),
	\label{eq:single-photon exponential wavepacket}
	\end{equation}
	is sent in, with the 2LS initially excited, so $\psi(x,0)=\varphi(x)$ and $\chi(x_1, x_2, 0)=0$. In this case, one also needs to specify the wavepacket width $\alpha$ in the input file. 
	In $x<-a$, the boundary condition is given by 
	\begin{equation}
	\psi(x,t) = \varphi(x-t)\times e_1(t),
	\end{equation}
	where $e_1(t)$ is the 2LS wavefunction solved in the one-excitation sector with initial conditions $e_1(0)=1$ and $\phi(x, 0) =0$,
	\begin{equation}
	e_1(t)=e^{-(i\omega_0+\frac{\Gamma}{2})t}\sum_{n=0}^{\infty}\frac{1}{n!}\left[\frac{\Gamma}{2}e^{(i\omega_0+\frac{\Gamma}{2})2a}(t-2na)\right]^n\theta(t-2na).
	\label{eq:spontaneous emission solution}
	\end{equation}
	
	\item[(c)] \emph{Two-photon exponential wavepacket} \cite{FrancescoBIC}: 
	The corresponding initial conditions are $\psi(x,0)=0$ and
	\begin{equation}
	\chi(x_1, x_2, 0) = \frac{A}{\sqrt{2}}\left[\varphi_1(x_1)\varphi_2(x_2)+\varphi_1(x_2)\varphi_2(x_1)\right],
	\end{equation} 
	where $A$ is the normalization constant such that $\iint dx_1 dx_2 |\chi(x_1, x_2, 0)|^2=1$ and $\varphi_i(x)$ is the $i$-th photonic wavepacket, assumed of the form Eq.~\eqref{eq:single-photon exponential wavepacket} with incident frequency $k_i$ and width $\alpha_i$.
	The normalization constant can be chosen to be positive without loss of generality:
	\begin{equation}
	A = \sqrt{\frac{4(k_1- k_2)^2+(\alpha_1+\alpha_2)^2\Gamma ^2}{4(k_1- k_2)^2 +(\alpha_1+\alpha_2)^2\Gamma ^2+4\alpha_1\alpha_2\Gamma ^2}}.
	\end{equation}
	
	Following our standard procedure, we first solve for $\psi(x<-a, t)$ and then plug it into the FDTD code to solve in the region $x>-a$. We find that the solution is given by 
	\begin{equation}
	\psi(x<-a, t) = A \left[\varphi_1(x-t) e_0^{(2)}(t) + \varphi_2(x-t) e_0^{(1)}(t) \right],
	\end{equation}
	where $e_0^{(i)}(t)$ is the qubit wavefunction in the one-excitation sector, solved subject to $e(0)=0$ and $\phi(x,0)=\varphi_i(x)$:
	\begin{align}
	e(t)&=\frac{\sqrt{\alpha\Gamma^2/2} (e^{-(i\omega_0+\Gamma /2)t}-e^{-(ik+\alpha\Gamma/2)t})}{p} 
	-i\sqrt{\alpha\Gamma} \sum_{n=1}^{\infty}\frac{\left(\frac{\Gamma}{2}\right)^{n-1/2}}{n!}
	\Bigl[(t-2na)^n e^{-(i\omega_0+\Gamma/2)(t-2na)}\nonumber \\
	&\quad\quad+\frac{i^n(k-\omega_0-i\alpha\Gamma/2)}{p^{n+1}}\gamma(n+1,-ip(t-2na))e^{-(ik+\alpha\Gamma/2)(t-2na)}
	\Bigr]\theta(t-2na)
	\label{eq:exact solution single-photon exponential}
	\end{align}
	with $p=k-\omega_0+i\Gamma/2(1-\alpha)$.
	Note that when the two photons are identical, $\varphi_1=\varphi_2=\varphi$, the expression above reduces to the known form: $\psi(x<-a, t) = \sqrt{2}\varphi(x-t)e_0(t)$ [cf.\ Eq.~\eqref{eq: BC two-photon scattering}]. 
\end{enumerate}
One could easily tweak the code to accept other kinds of initial conditions, but the strategy for solving Eq.~\eqref{eq:double-excitation delay differential eq} remains the same for all possible scenarios. It is important to note that due to the nature of the delay PDE \eqref{eq:double-excitation delay differential eq}, one cannot solve it numerically without knowing its analytical solutions in $x<-a$. But one do not need more than that either --- the constraint $n_x\leq 2N_x$ guarantees that the knowledge in $x<-a$ is sufficient, as it makes the line $x=-a$ lie outside of the green region in Fig.~\ref{fig:FDTD_schematic} so that the boundary condition is completely determined. 

Finally, to calculate various non-Markovian measures using the solution of $\psi$, instead of post-processing it is much easier to do most of the work \emph{in situ}. To construct the geometric measure \cite{LorenzoPRA13}, two functions $\lambda(t)$ and $\mu(t)$ can be calculated \cite{FangNJP18}.\footnote{In Ref.~\cite{FangNJP18} $\mu(t)$ is called $c(t)$ and $\lambda(t)$ is called $\Delta(t)$, respectively. The author regrets the inconvenience.\label{note5}}
The detail of the implementation is presented in \ref{append: NM}. 

\section{Parallelization}
As discussed above, due to the non-local delay term there is a strong data dependency --- the present solution depends on its past history. However, even with such a severely constrained problem, interestingly we find that the delay PDE \eqref{eq:double-excitation delay differential eq} can still be solved in parallel, by which we mean that the simultaneous presence of multiple FDTD solvers in the spacetime is allowed.
In this paper, we propose and implement two kinds of parallel approaches, referred to as the ``swarm'' and ``wavefront'' approaches, respectively, that respect the constraints imposed by the delay PDE \eqref{eq:double-excitation delay differential eq}. 

\begin{figure}[hbtp]
	\centering
	\includegraphics[width=0.7\textwidth]{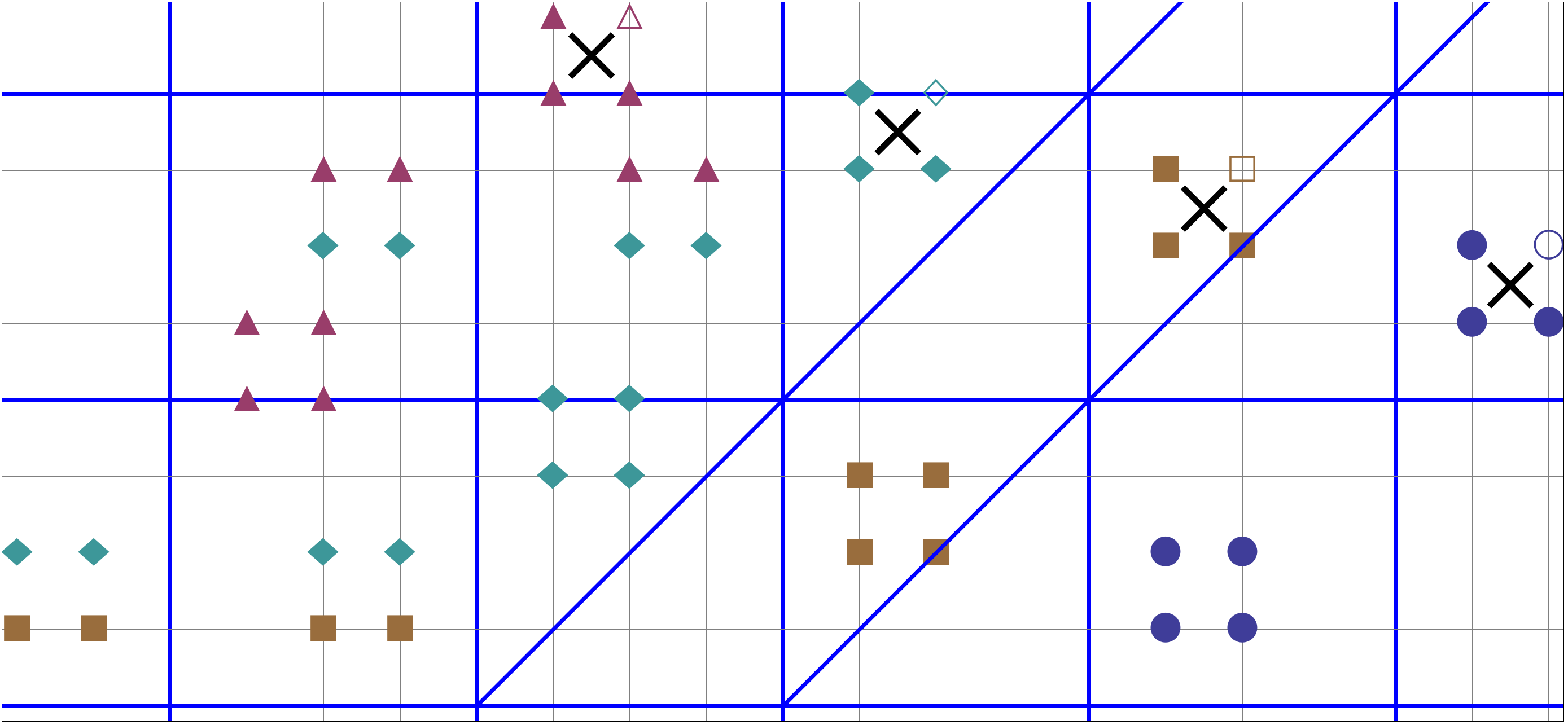}
	\caption{A snapshot of multiple marching FDTD solvers in the swarm approach. Each color and shape corresponds to points read and written by a certain solver (thread). The points to be solved by each solver are empty, while known points are filled. Note that each solver is one step above and at least $n_x$ steps behind its predecessor (here $n_x=4$).}
	\label{fig: FDTD multithread marching}
\end{figure}

For the swarm approach (see Fig.~\ref{fig: FDTD multithread marching}), when solving Eq.~\eqref{eq:double-excitation delay differential eq} from the bottom up,
each solver must be one temporal step above and at least $n_x$ spatial steps behind its next neighbor (called predecessor hereafter), or the execution order would not be preserved correctly; that is, one solver may read the value at certain point which is yet to be solved by another solver. It turns out that it is beneficial to have a ``cyclic lag'' relation among the solvers: denoting the total number of solvers as \texttt{Nth}, which can be set in the input file, then \#2 is lagged behind \#1, \#3 behind \#2, etc., and finally \#1 can be set behind \#\texttt{Nth} once it reaches the grid boundary. We find that such a wrap-around can increase performance and is easy for coding. 

In the current implementation, we utilize the ``wait-and-signal'' mechanism provided by POSIX Threads (pthreads). Each solver is marched by an independent thread and has a lock for its position and a condition variable for waiting for its predecessor. Before attempting to solve at a certain point, it must look up its predecessor's position and see whether the constraint is satisfied or not. If not, then it must wait for the signal sent by its predecessor before proceeding. 
Using pthreads' locks not only allows a solver to communicate with and wait for its predecessor, but also preserves cache coherence, which is important since very often a solver needs to access a value immediately after it's solved by another solver. 

\begin{figure}[hbtp]
	\centering
	\includegraphics[width=0.7\textwidth]{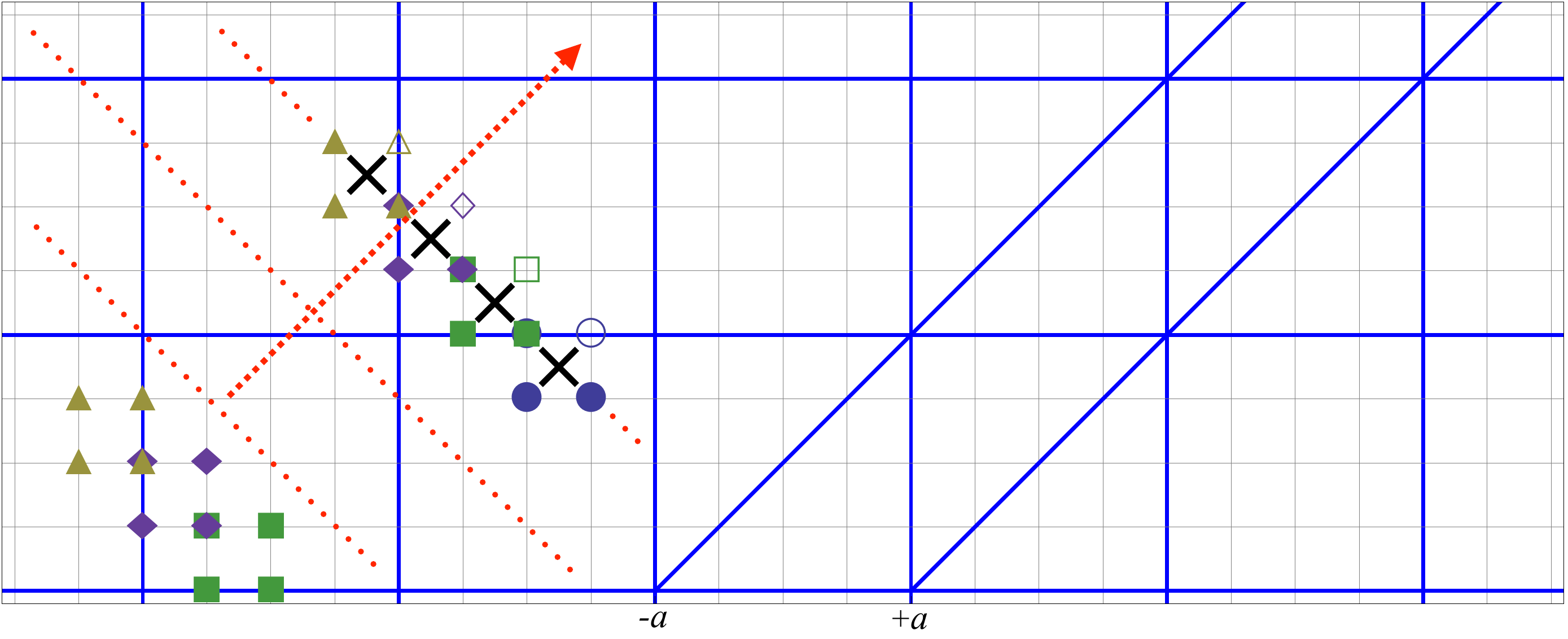}\\
	\includegraphics[width=0.7\textwidth]{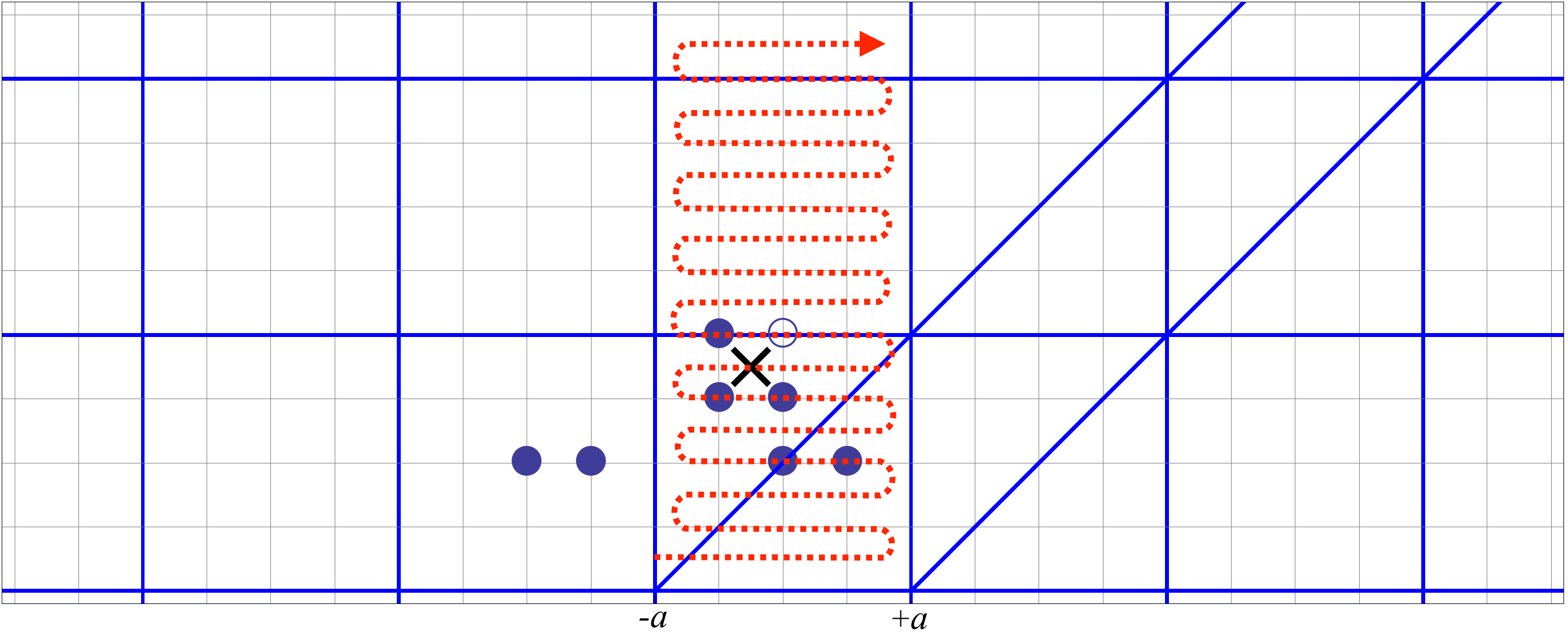}\\
	\includegraphics[width=0.7\textwidth]{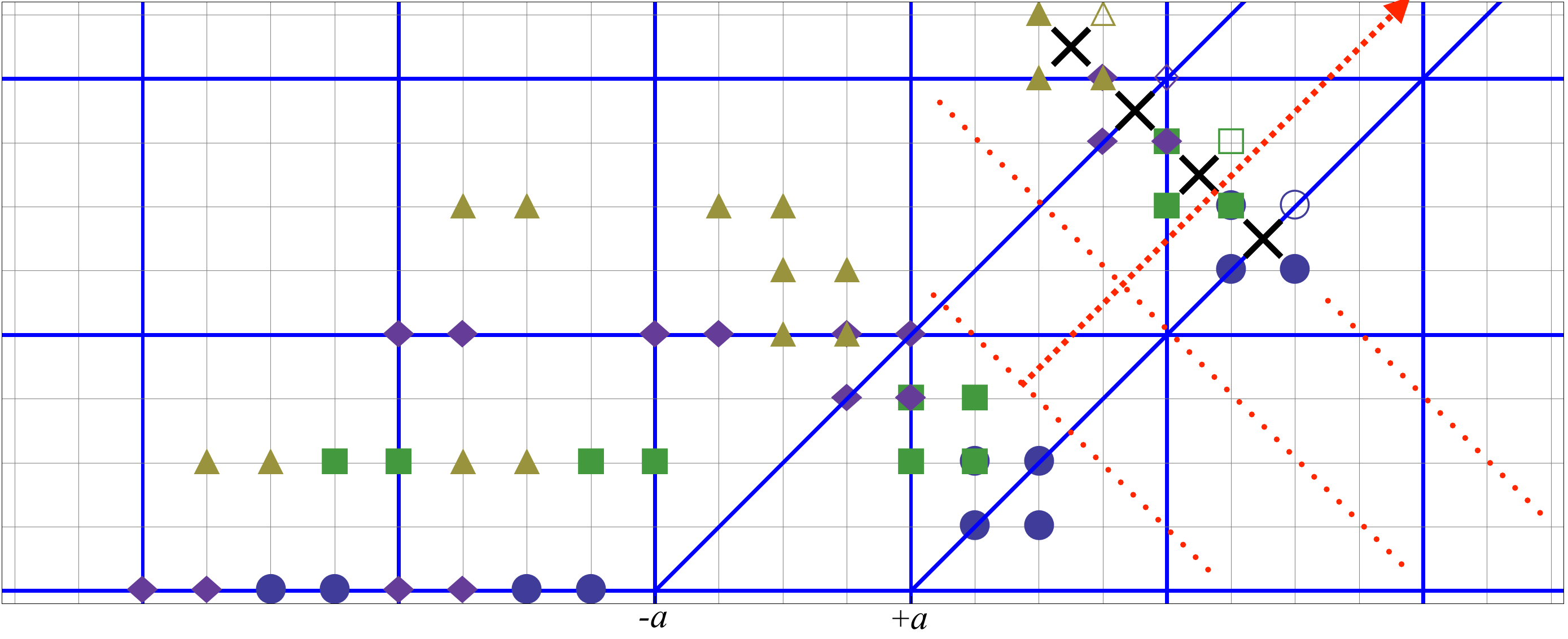}
	\caption{Schematic of the wavefront  approach. We first solve in the region $x\leq -a$ (top), then $-a<x\leq a$ (middle), and finally $x>a$ (bottom). In the first and last regions, points on the same wavefront (represented by dotted lines) can be solved in parallel. In the middle region, due to causality only one solver is permitted to enter and move row by row. The solver(s) are propagated along the dashed lines in all regions. The colors and shapes of all points have the same meaning as in Fig.~\ref{fig: FDTD multithread marching}.}
	\label{fig: FDTD wavefront marching}
\end{figure}

For the wavefront approach (see Fig.~\ref{fig: FDTD wavefront marching}), we circumvent the requirement of separating $n_x$ steps in space by partitioning the spacetime into three regions: $x\leq-a$, $-a<x\leq a$, and $x>a$, which are solved in turn. In the first and last regions, Eq.~\eqref{eq:double-excitation delay differential eq} can be solved by propagating the wavefront along the diagonal, and points on the same wavefront (along which the spatial lag between adjacent solvers is only one step) can be solved in parallel. In the middle region $-a<x\leq a$, which is $n_x$-step wide, due to causality  
we use only one solver (as if no parallelization exists). Physically, it is clear \cite{FangNJP18} that in $x\leq-a$ ($x>a$) we only have right- (left-) going photons, and in $-a<x\leq a$ they are connected by the mirror.

We implement the wavefront approach using OpenMP, since within each wavefront it is simply an \emph{embarrassingly parallel} problem; that is, there is no data dependency and thus no need to lock or signal. Moreover, because most of the time a wavefront typically contains at least hundreds of points, the synchronization penalty (due to the implicit barrier in the OpenMP constructs) is small compared to the gain from parallelization, and the barrier in turn helps preserve the cache coherence.

In the end of the next section, we present a strong-scaling measurement for a typical problem size required by the physics \cite{FangNJP18,FangPRA15err,FrancescoBIC}. It will be shown that the wavefront approach outperforms the swarm approach.

\section{Validation}
User instruction for the program is given in the \texttt{README.md} file distributed along with the code.
We have tested our program on both Linux and Mac OS X.
There are at least three possible tests to pass for establishing the validity of this code: 1. $\psi$ in $x\leq-a$, where the full, exact solution can be obtained; 2. $\psi$ in $-a<x\leq a$ because we can solve for the first few triangular tiles analytically \cite{FangNJP18}; 3. the two-photon wavefunction $\chi$, because the steady-state result is known from scattering theory \cite{FangPRA15}.

\begin{figure}[htbp]
	\centering
	\includegraphics[scale=0.5]{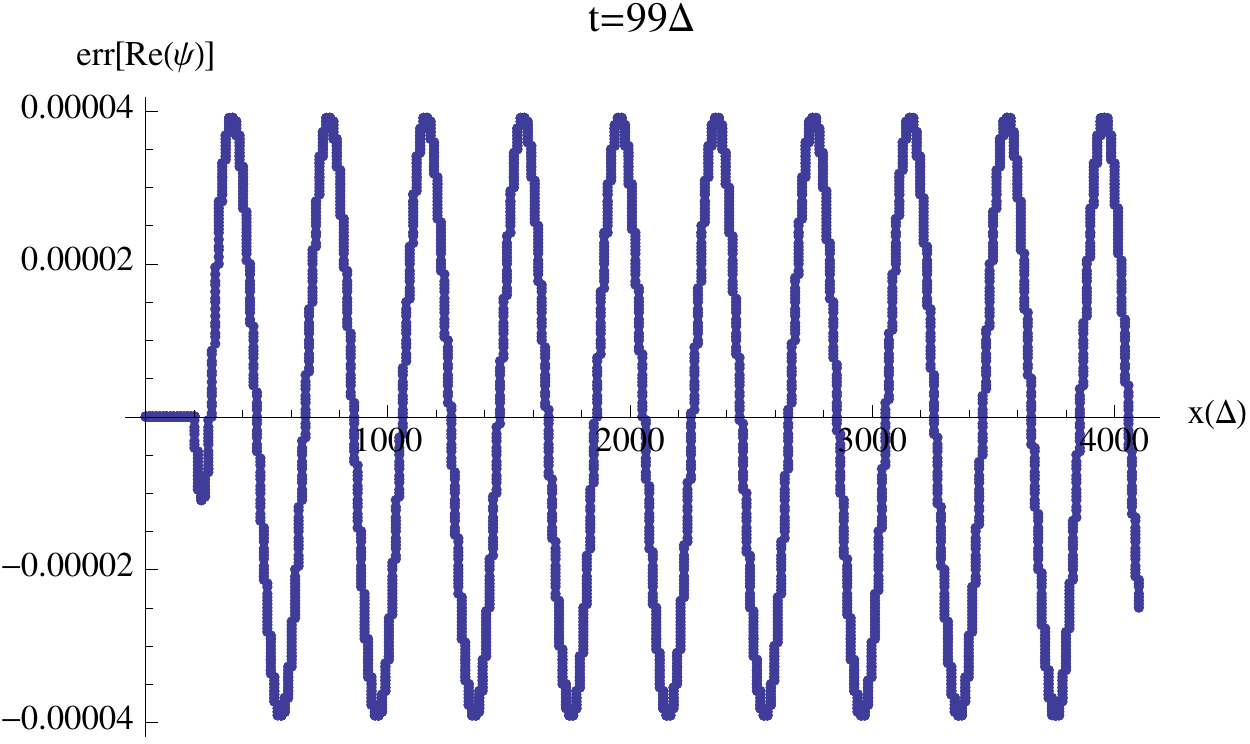}
	\includegraphics[scale=0.5]{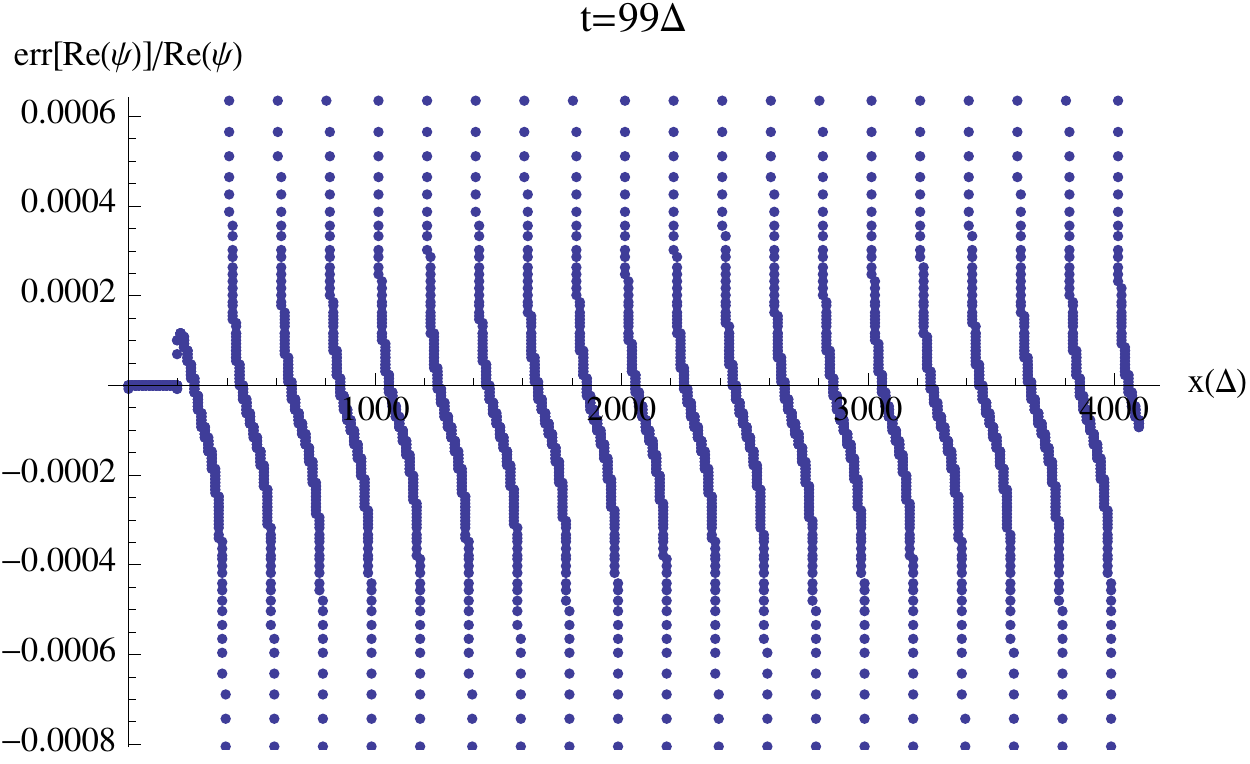}\\
	\vspace{0.5cm}
	\includegraphics[scale=0.5]{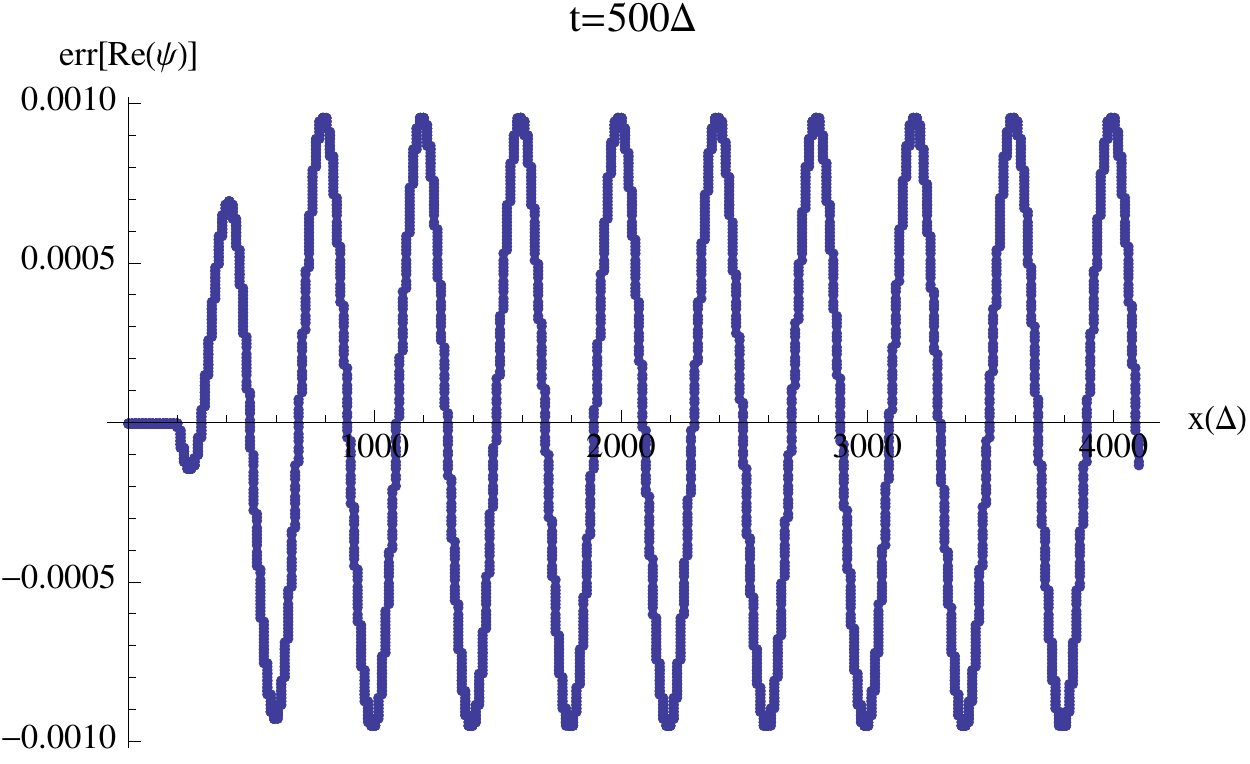}
	\includegraphics[scale=0.5]{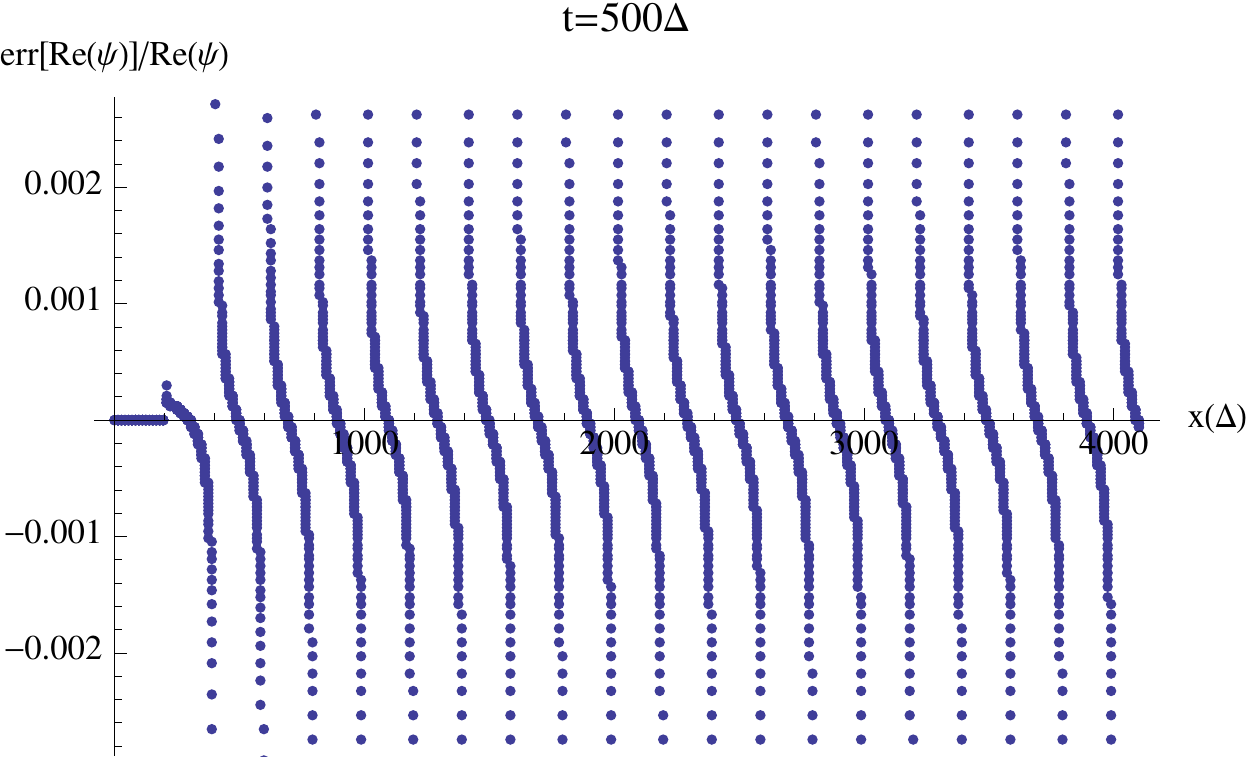}\\
	\vspace{0.5cm}
	\includegraphics[scale=0.5]{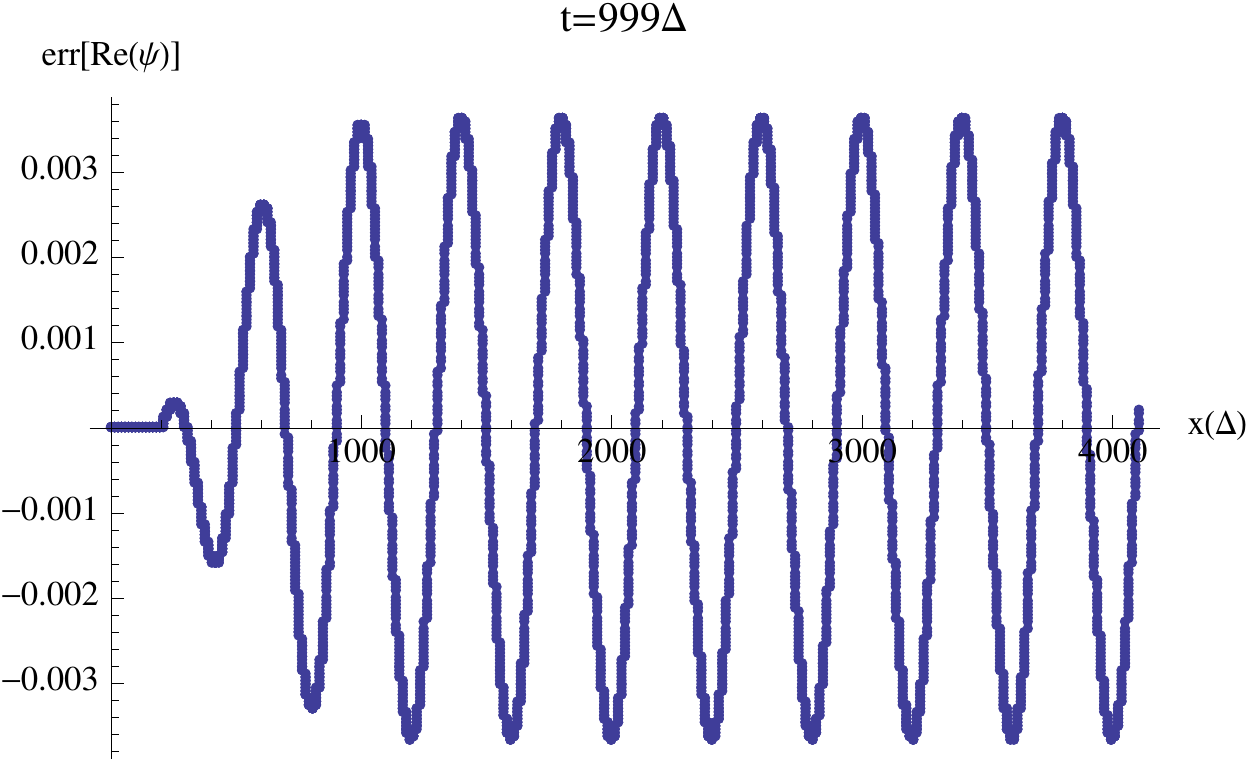}
	\includegraphics[scale=0.5]{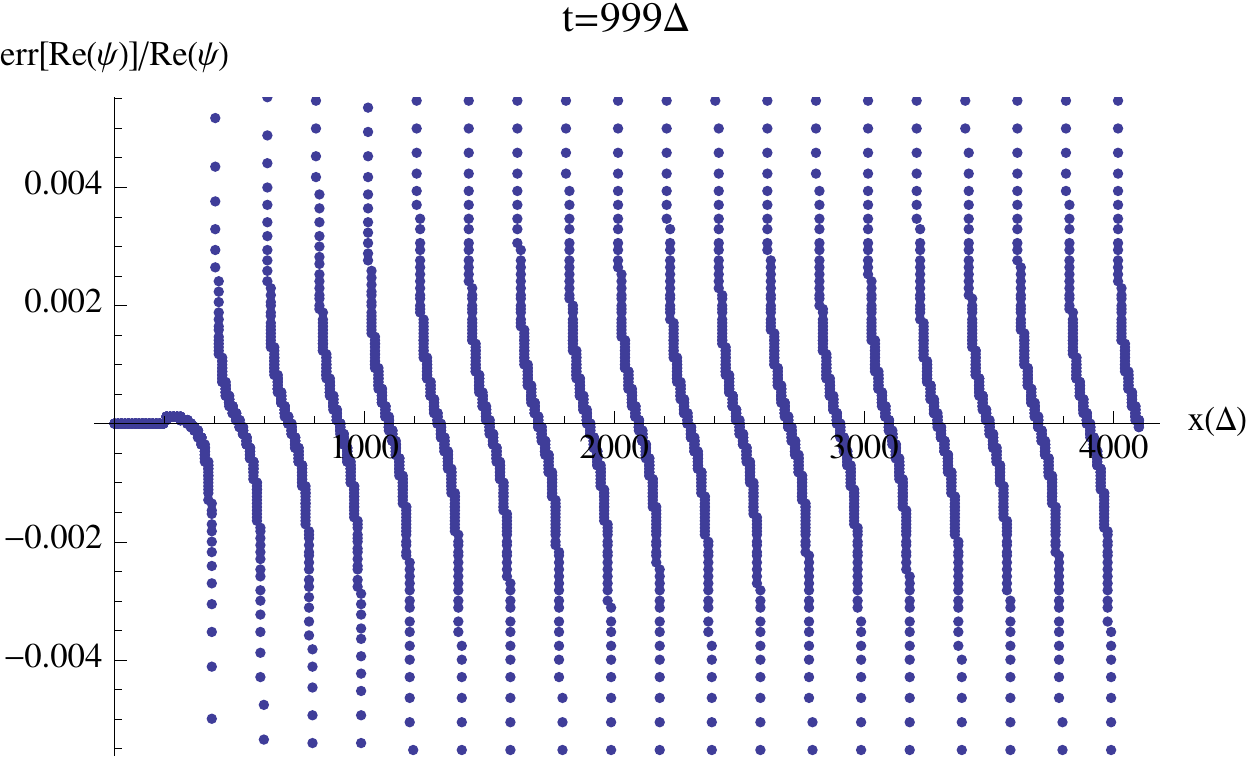}\\
	\vspace{0.5cm}
	\includegraphics[scale=0.5]{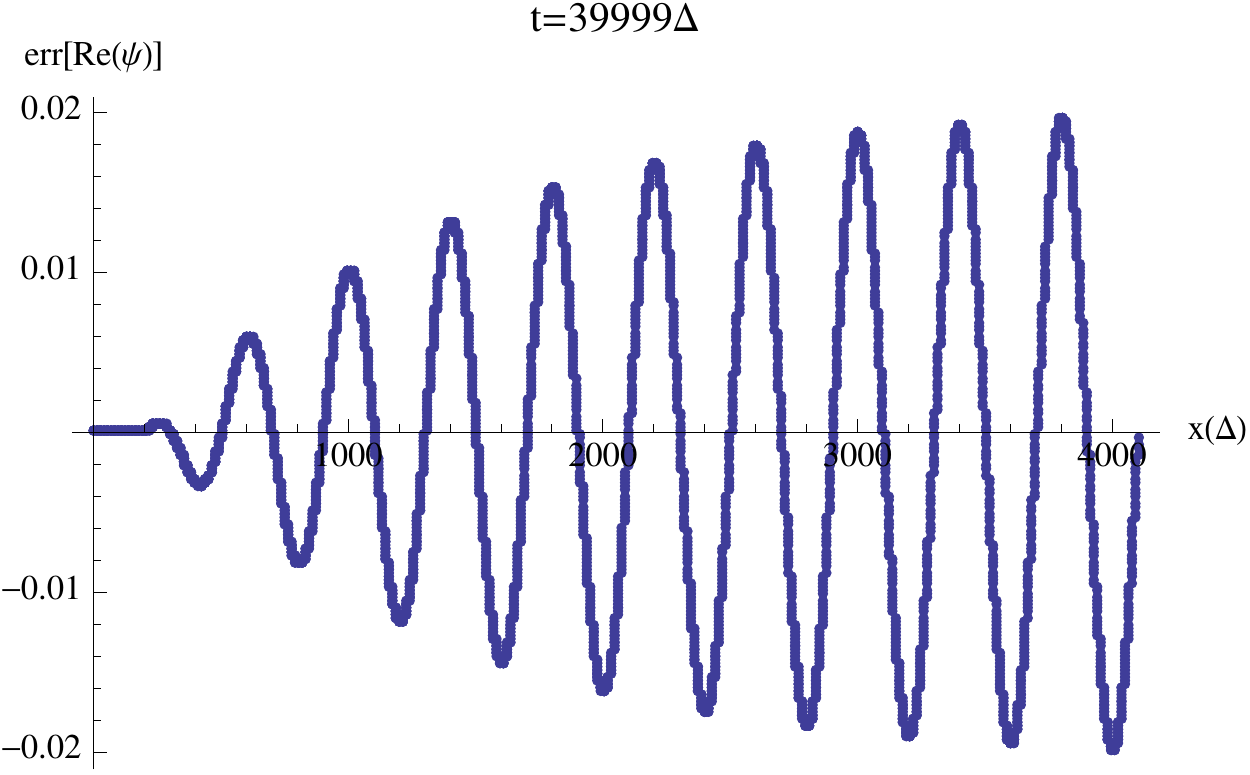}
	\includegraphics[scale=0.5]{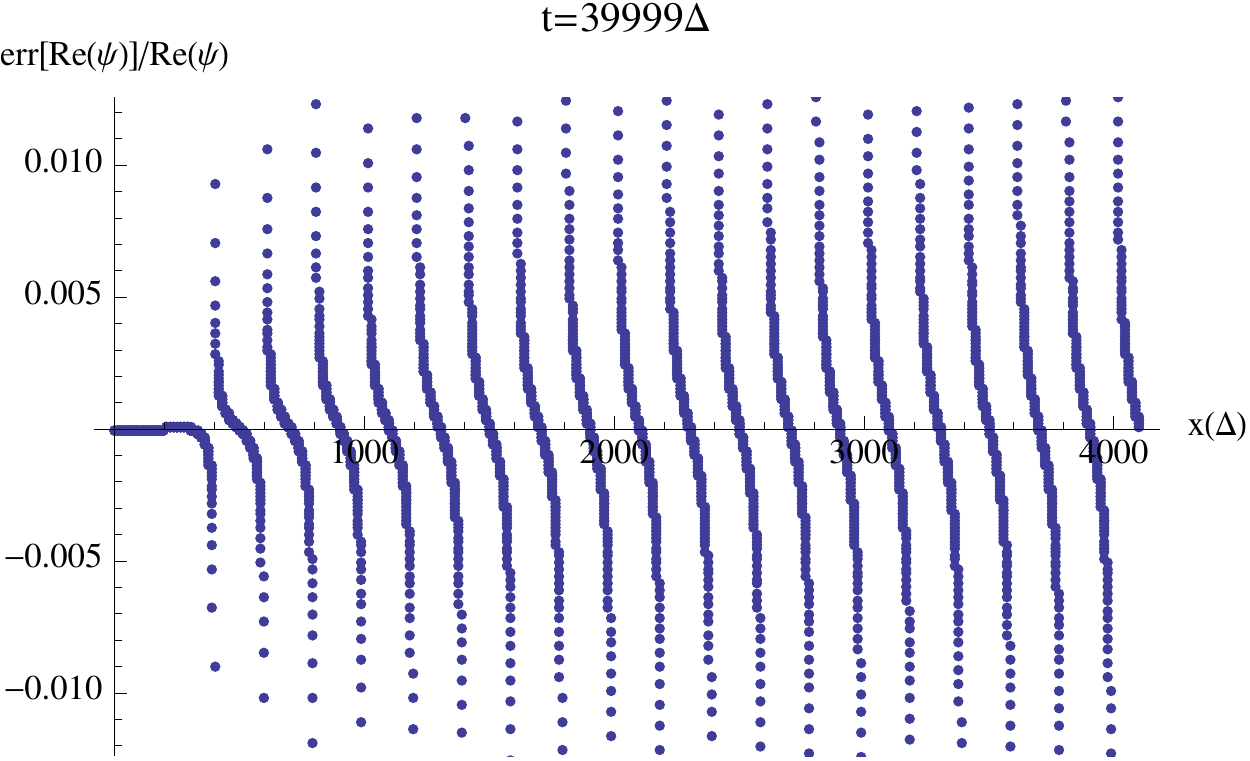}
	\caption{Absolute (left) and relative (right) errors for $\re(\psi)$ as a function of spatial steps (in units of $\Delta$) at $t/\Delta=\{99, 500, 999, 39999\}$ (from top to bottom). The last step corresponds to $x=-a$. Input parameters: $n_x=200, N_x=4000, N_y=4\times10^4, \Delta=10^{-2}, k=\omega_0=\pi/2, \Gamma=\pi/40$.}
	\label{fig:error}
\end{figure}

For the first test, in Fig.~\ref{fig:error} we show several snapshots of the absolute and relative errors of the real part of $\psi$ as a function of steps in the $x$-direction (and $x\leq-a$ in all plots).
We note that in order to reach the steady state, usually the termination time needs to satisfy $t\gg1/\Gamma$. 
In addition, note two common features in these plots: (a) the error is zero in the initial steps because it's where the boundary condition resides; (b) errors are well-controlled in the sense that they are bounded in an oscillating envelope, so the program behaves correctly in $x\leq-a$.

%
%
%

\begin{figure}[!t]
	\centering
	\includegraphics[scale=0.5]{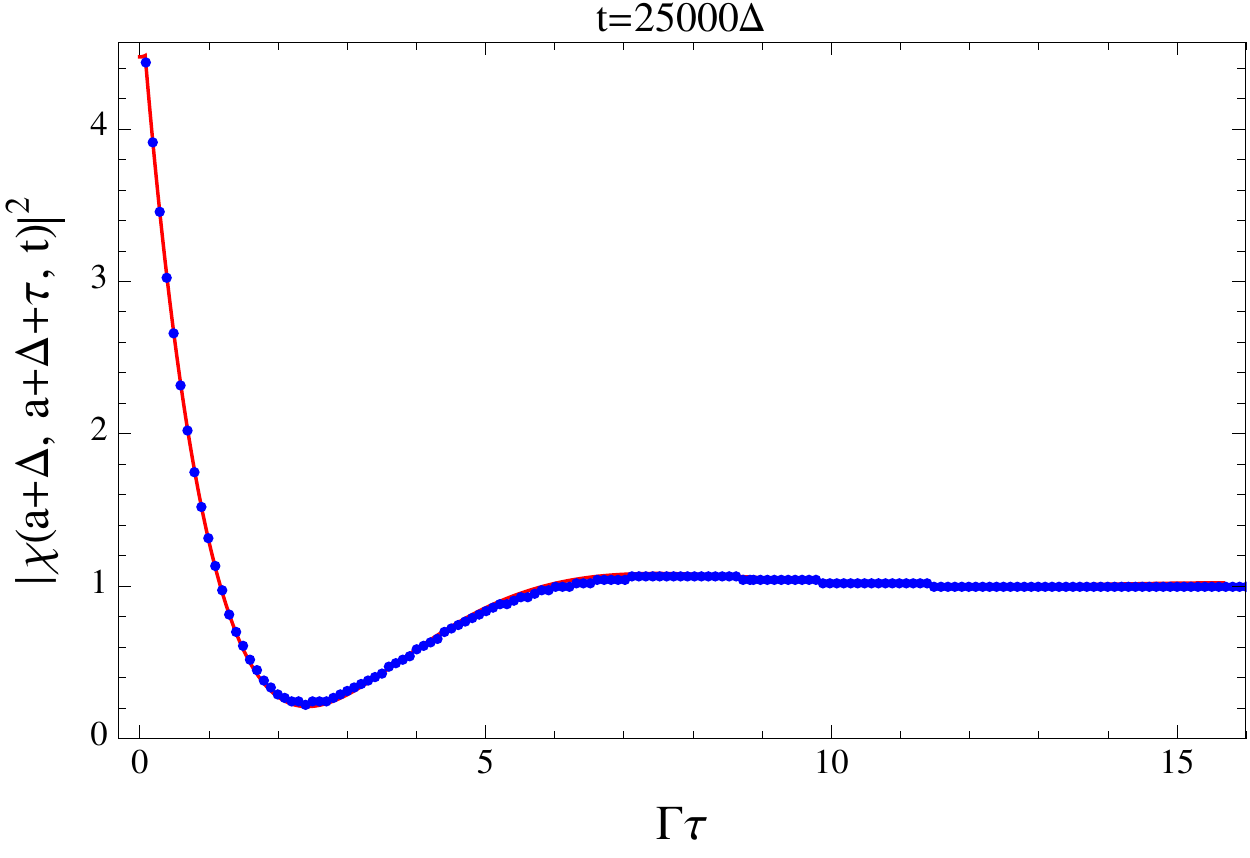}
	\includegraphics[scale=0.5]{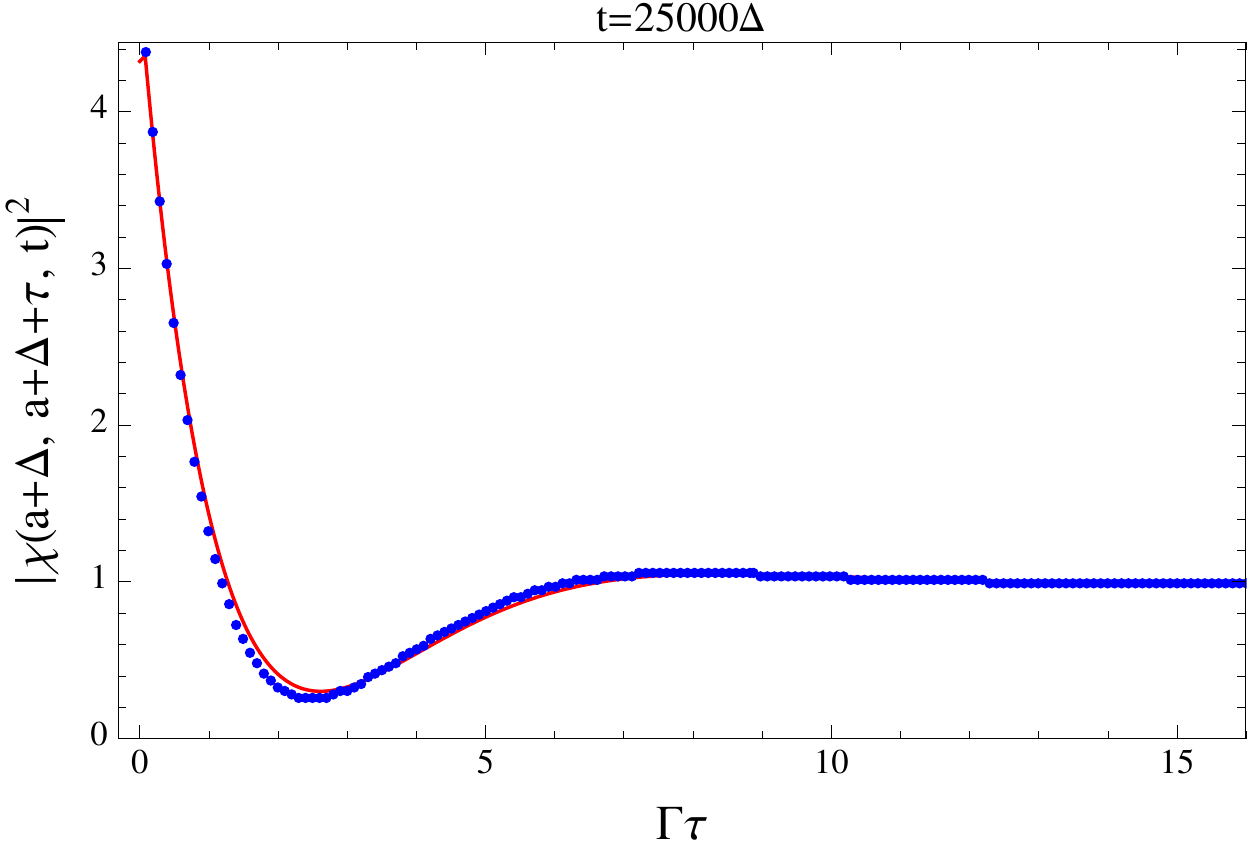}
	\caption{Comparison of calculated two-photon wavefunction $|\chi|^2$ (red curves) as a function of photon separation $\uptau$ in the long-time limit for $k_0a=\pi/4$ and (left) $k=\omega_0$ (right) $k=\omega_0-\Gamma$. The blue dots are from the numerically exact scattering theory \cite{FangPRA15}. Input parameters: $n_x=50, N_x=10^4, N_y=2.5\times10^4, \Delta=10^{-2}, \omega_0=3.1415926536, \Gamma=0.1570796327$.}
	\label{fig:g2 piover4}
\end{figure}

\begin{figure}[!t]
	\centering
	\includegraphics[scale=0.5]{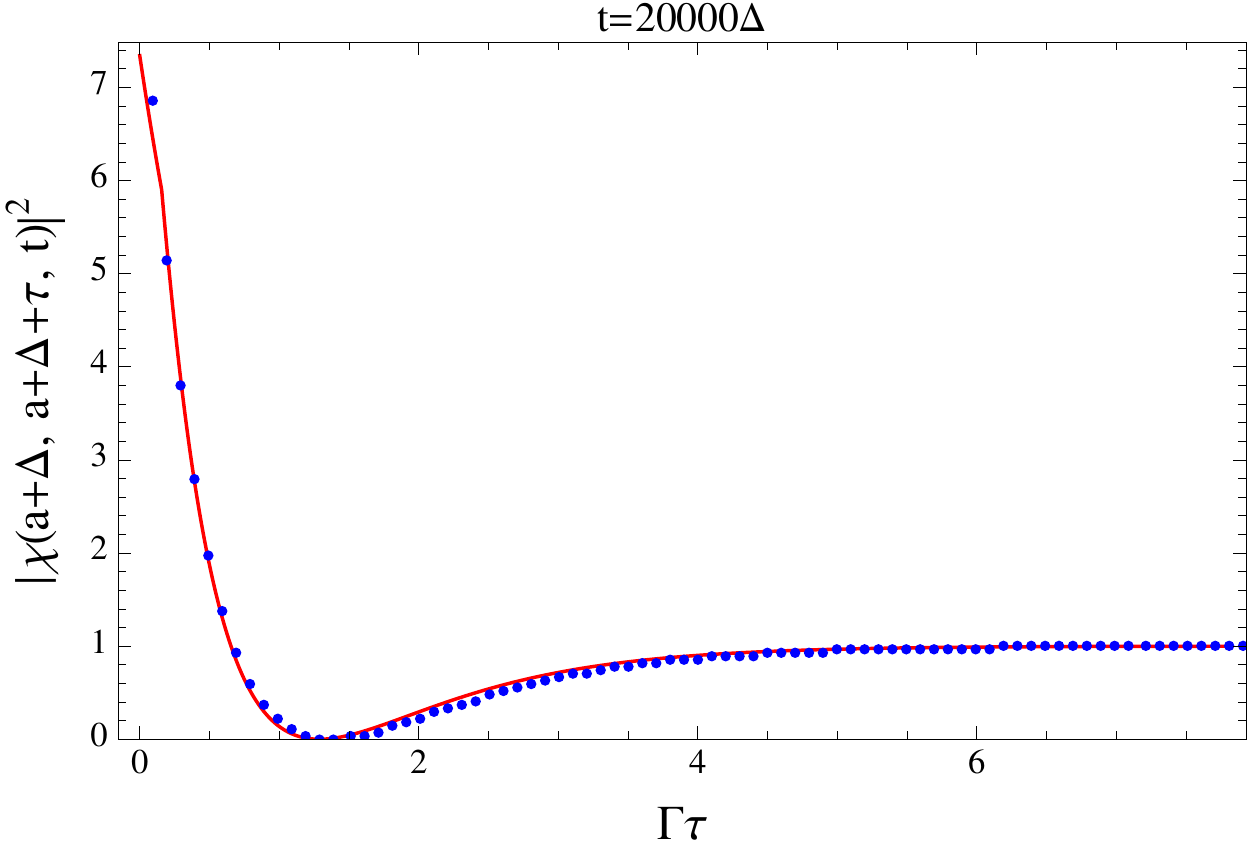}
	\includegraphics[scale=0.5]{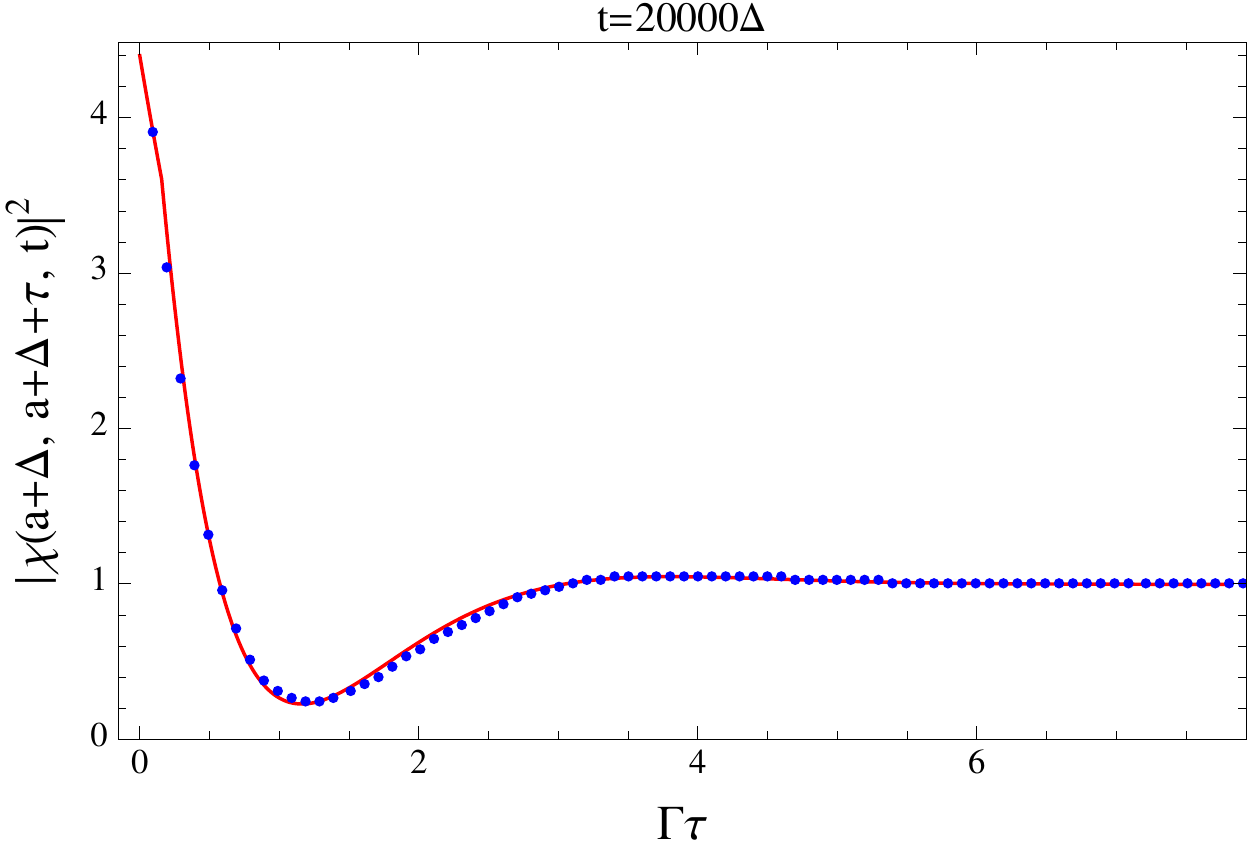}
	\caption{Comparison of calculated two-photon wavefunction $|\chi|^2$ (red curves) as a function of photon separation $\uptau$ in the long-time limit for $k_0a=\pi/2$ and (left) $k=\omega_0$ (right) $k=\omega_0-\Gamma$. The blue dots are from the numerically exact scattering theory \cite{FangPRA15}. 
		Input parameters: $n_x=200, N_x=10^4, N_y=2\times10^4, \Delta=10^{-2}, \omega_0=1.5707963268, \Gamma=0.0785398163$.}
	\label{fig:g2 piover2}
\end{figure}

\begin{figure}[!t]
	\centering
	\includegraphics[scale=0.5]{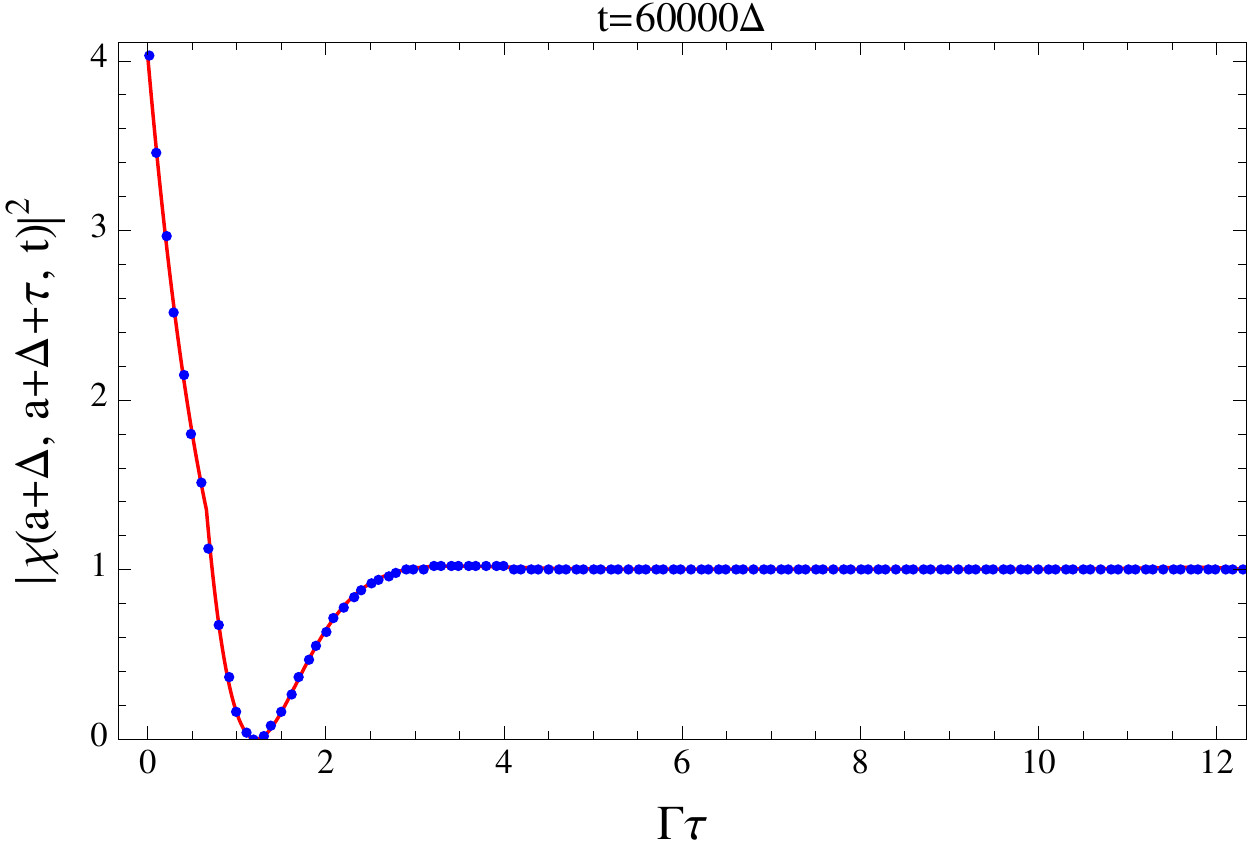}
	\includegraphics[scale=0.5]{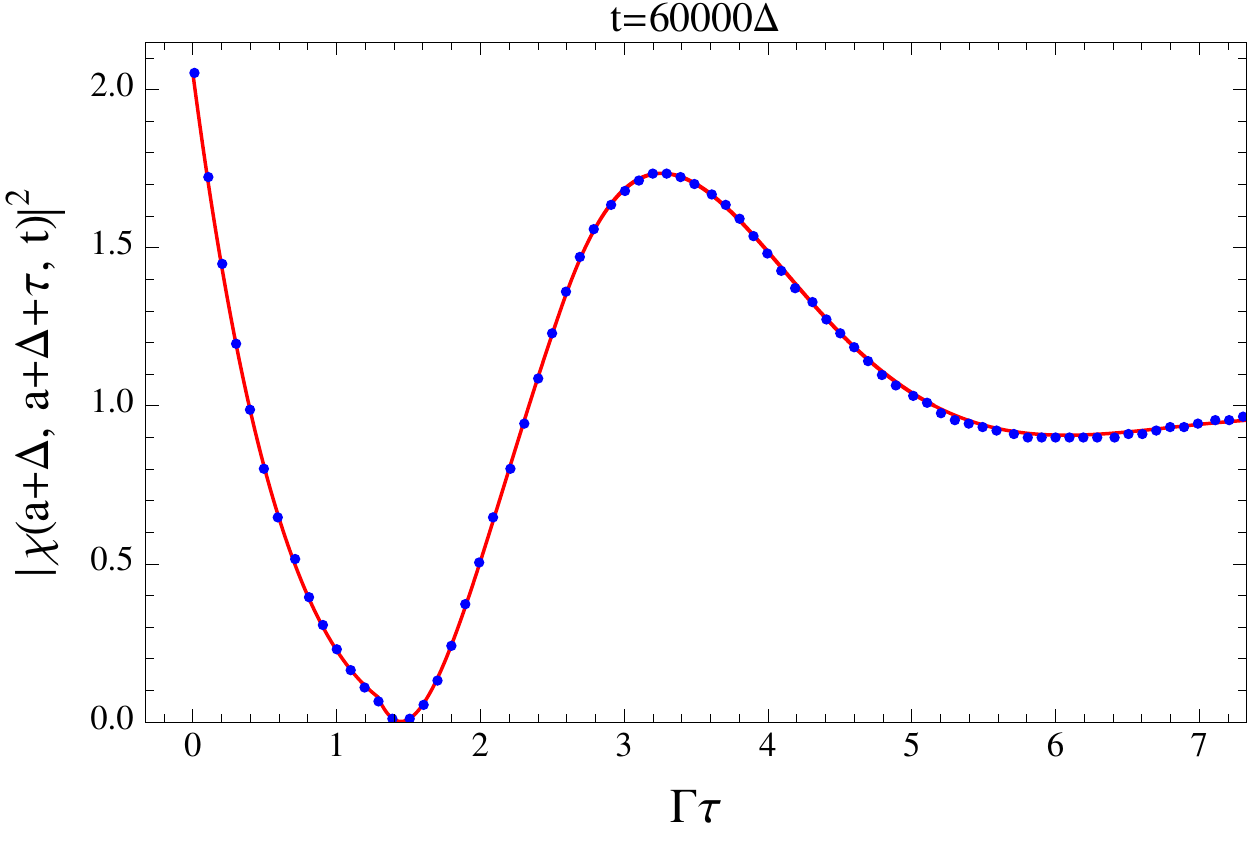}
	\caption{Comparison of calculated two-photon wavefunction $|\chi|^2$ (red curves) as a function of photon separation $\uptau$ in the long-time limit for $k=\omega_0$ and (left) $k_0a=10.5\pi$ (right) $k_0a=20.5\pi$. The blue dots are from the numerically exact scattering theory \cite{FangPRA15}. 
		Input parameters: $N_x=6.48\times10^4, N_y=6\times10^4, \Delta=\pi/12000, k=\omega_0=100, \Gamma=1$.}
	\label{fig:g2 NM}
\end{figure}


For the second test, we generated the analytical expressions of $\psi$ in the first four tiles (the fifth one took too much time to compute) in Mathematica \cite{FangNJP18} and then compared the values on the FDTD grid points. The numerical result agreed well too (not shown).

Finally, to compare with the scattering theory we can construct the two-photon wavefunction $\chi(x_1, x_2, t)$ using the calculated $\psi$ according to the formal solution of $\chi$, Eq.~\eqref{eq:formal solution of two-photon wavefunction},
where the coordinates $x_1$ and $x_2$ should be chosen as $x_{1,2}\geq +a$ for capturing the outgoing fields. In the program, $\chi$ is calculated by setting $x_1=a+\Delta$ and $x_2=a+\Delta+\uptau$, with $\uptau=x_2-x_1$ being the separation of the two detectors.\footnote{We note that by definition the two-photon correlation function is given by $g_2(\uptau) = |\chi(x_d, x_d+\uptau, t\rightarrow\infty)|^2$ with $x_d\geq+a$ ($x_d+\uptau\geq+a$) being the position of the first (second) detector.} The results are shown in Figs.~\ref{fig:g2 piover4}-\ref{fig:g2 NM}. The agreement is quite well, even in the non-Markovian regime in which $\Gamma a/c\gtrsim1/2$ (Fig.~\ref{fig:g2 NM}). Because evaluating Eq.~\eqref{eq:formal solution of two-photon wavefunction} requires the full history of $\psi$, it is clear the program is valid for all regions in the spacetime.
More results generated by our FDTD program are discussed in Refs.~\cite{FangNJP18,FangPRA15err,FrancescoBIC}.

\begin{figure}[htbp]
	\centering
	\includegraphics[width=0.7\textwidth]{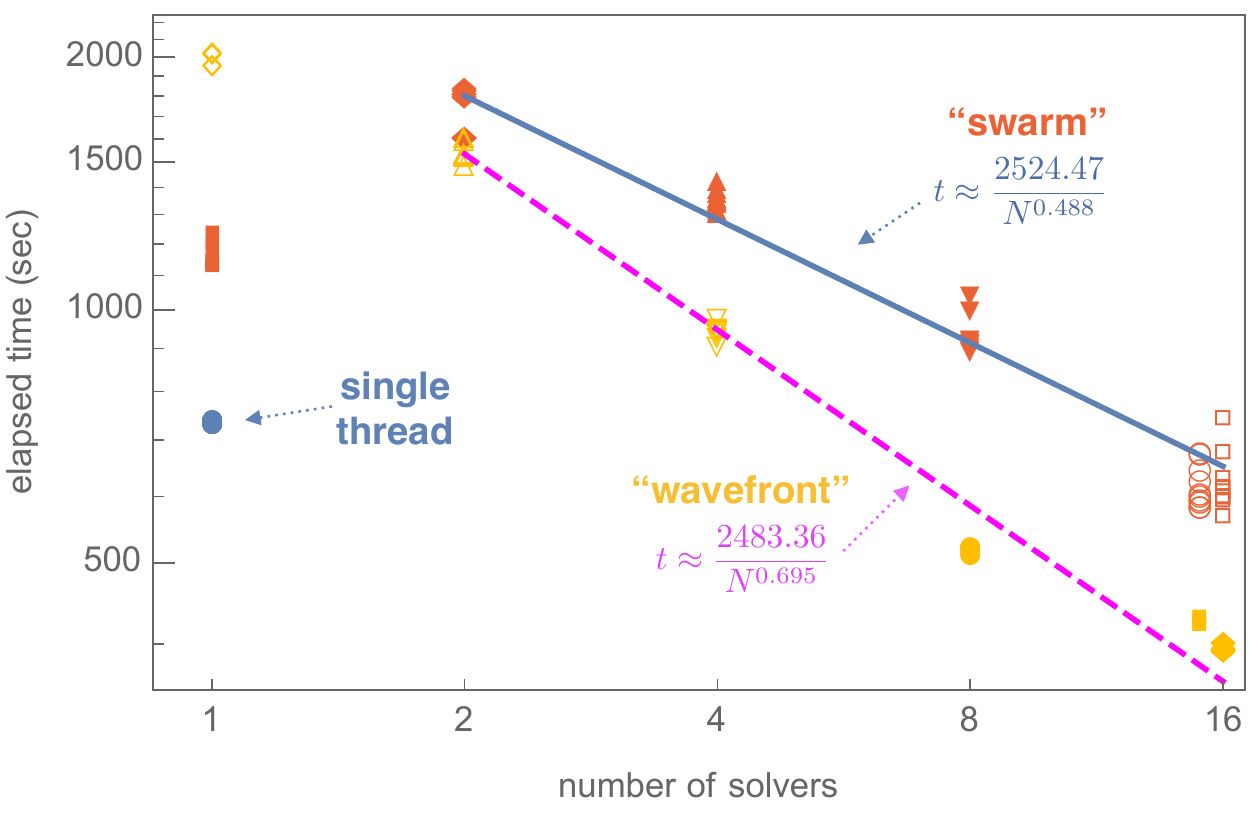}
	\caption{Log-log plot of measured elapsed time as a function of the number of solvers (threads) set by \texttt{Nth} using the swarm (orange) and the wavefront (yellow) approaches. The test is run 10 times for each case. For single thread, the measurements without threading overhead (light blue circles) are also shown. The straight lines are fitted by all data points with $\texttt{Nth}=1$ excluded. Input parameters: $N_x=64800$, $N_y=60000$, $n_x=960$, $\Delta=2.618\times10^{-4}$, \texttt{init\_cond}$=2$, $k=\omega_0=100$, $\Gamma=5$, and $\alpha=0.5$. Test environment: Intel Xeon CPU E5-2670 \@ 2.6 GHz with 16 physical cores, 128 GB physical memory, and gcc 7.2.0.}
	\label{fig: elapsed time for multithread}
\end{figure}

We next comment briefly on the performance of multi-thread support. In Fig.~\ref{fig: elapsed time for multithread} we report the elapsed time as a function of the number of threads (specified by \texttt{Nth}) for solving Eq.~\eqref{eq:double-excitation delay differential eq} using both the swarm and wavefront approaches for the same input parameters, which are chosen to roughly match those used in Refs.~\cite{FangNJP18,FangPRA15err,FrancescoBIC}. 
In order to have a fair comparison, note that when single thread is used, the threading libraries would impose unnecessary overhead, which is enormous in particular for the wavefront approach since after every step a barrier synchronization is enforced. Therefore, we also report in Fig.~\ref{fig: elapsed time for multithread} the measurement without using any threading mechanism (light blue circles), compared with which the speedup is close to 2x using the wavefront approach with 16 threads. We find that the wavefront approach performs much better than its swarm counterpart: it has a better scaling and is already faster than single thread with only 8 threads. However, both approaches do provide shorter runtime than that of the serial version. Finally, we note in passing that the test is performed on a machine with hyperthreading turned off, so we are unable to test with more threads, but we expect to see a consistent scaling behavior when running on machines with $>16$ physical or logical cores.

\section{Conclusion}
The program provides a numerically exact, \emph{time-dependent} solution to the problem of a single 2LS placed in front of a mirror scattered by either a one- or two-photon initial state, so for waveguide-QED researchers this program can be very useful for solving the three classes of problems as presented above. Arbitrary wavepackets or other forms of initial conditions can be incorporated into the code with minor modification. 

More flexibility, such as a giant atom that has arbitrary number of distant legs coupled to the waveguide \cite{GustafssonSci14,FriskKockumPRA14,GuoPRA17}, can be gained by re-implementing this program in an OO language (C++, Python, etc). In fact, the code has been written in the OO-style by (i) compacting all relevant fields into a C struct (roughly equivalent to a C++/Python class) named \texttt{grid}, (ii) passing ``this'' pointer to a \texttt{grid} instance when calling functions, as if they were member functions of the \texttt{grid} class, and (iii) following RAII. As a result, rewriting in C++, for example, should be of minimal work. Of course, the grid layout must be carefully designed in such a scenario to minimize memory usage.
As for the case of an infinite waveguide, we believe the best design is to use MPI for communication between the left- and right-moving solvers, which we leave for future work.

Finally, this program serves as a proof of concept for mathematicians, scientists and engineers who are interested in solving complicated delay PDE using parallel FDTD. Important issues such as general proof of stability, more efficient multi-thread implementation, robust optimization over any 1+1D delay PDE, etc., remain challenging. For development discussions, please either open an issue on GitHub\footnote{\url{https://github.com/leofang/FDTD}} or contact the authors. Users and developers who are benefited from this program are strongly encouraged to cite both this paper as well as Ref.~\cite{FangNJP18}.

\section{Acknowledgments}
We acknowledge financial support from U.S.\ NSF (Grant No.\ PHY-14-04125) and the Brookhaven National Laboratory's Laboratory Directed Research and Development project \#17-029, and fruitful discussion with Harold U.\ Baranger, Francesco Ciccarello and Meifeng Lin. We also thank Weiguo Yin for proofreading.
Part of the tests used resources of the Center for Functional Nanomaterials, which is a U.S.\ DOE Office of Science User Facility, at Brookhaven National Laboratory under Contract No. DE-SC0012704.

\appendix
\section{Stability analysis}
\label{appen: stability}
Following Ref.~\cite{NumericalRecipes}, we can perform a simple von Neumann stability analysis to check that our regularization scheme does not lead to amplitude divergence. First, let us consider the simplest ordinary PDE $\partial_x f+\partial_t f+W f(x,t)=0$ [compared with Eq.~\eqref{eq:double-excitation delay differential eq}, $W=\left(i\omega_0+\frac{\Gamma}{2}\right)$]. If we write $\psi(m\Delta, n\Delta)=\xi^n e^{ik(m\Delta)}$, where $\xi$ is called the amplification factor, then we have 
\begin{equation}
\xi(k) = \frac{\left(\frac{1}{\Delta}-\frac{W}{4}\right)-\frac{W}{4}e^{ik\Delta}}{\left(\frac{1}{\Delta}+\frac{W}{4}\right)+\frac{W}{4}e^{-ik\Delta}}e^{-ik\Delta}.
\end{equation}
It can be shown that $|\xi(k)|\leq 1$ as long as $\re(W)>0$, so our approach for solving this simple PDE is stable.

For Eq.~\eqref{eq:double-excitation delay differential eq} in $x<-a$, we can again insert the ansatz and obtain
(assuming $\chi=0$ for simplicity)
\begin{align}
\left(\frac{1}{\Delta}+\frac{W}{4}\right)\xi^n e^{ikm\Delta}
&=	\left(\frac{1}{\Delta}-\frac{W}{4}\right)\xi^{n-1} e^{ik(m-1)\Delta}
-\frac{W}{4}\left(\xi^{n-1}e^{ikm\Delta}+\xi^{n}e^{ik(m-1)\Delta}\right)\nonumber\\
&\quad+\frac{\Gamma}{8}\xi^{n-n_x-1}e^{ik(m-n_x-1)\Delta}\left(1+e^{ik\Delta}+\xi+\xi e^{ik\Delta}\right).
\end{align}
Note that the parameter $n_x$ enters because of the delay term. Now this is a polynomial in $\xi$ of order $n_x$, so there is no general solution for its roots. But empirically (i.e. by solving the equation graphically) we still find that $|\xi(k)|\leq 1$; that is, our algorithm is stable in $x<-a$.

A more general stability analysis for Eq.~\eqref{eq:double-excitation delay differential eq} is arduous. However, we hope the above reasoning together with the validations presented above are enough to convince the interested users that our FDTD program is stable.

\section{Conversions between physical quantities and simulation parameters} 
\label{appen: translation}
In the scattering problem, the three important parameters are $\mathcal{K}$ ($=k/\Gamma$), $\mathcal{W}$ ($=\omega_0/\Gamma$) and $n$ ($=k_0a/\pi$). To satisfy the rotating-wave approximation $\mathcal{W}\gg1$ is required, but how large $\mathcal{W}$ is should not matter. Once they are determined, the physics is determined. Here we describe how to express physical quantities in terms of $\Delta$, $n_x$, $\mathcal{K}$, $\mathcal{W}$ and $n$:
\begin{align}
k&=\frac{2n\pi\mathcal{K}}{n_x\mathcal{W}}\times\frac{1}{\Delta}\\
\omega_0&=\frac{2n\pi}{n_x}\times\frac{1}{\Delta}\\
\Gamma&=\frac{2n\pi}{n_x\mathcal{W}}\times\frac{1}{\Delta}\\
\lambda_0&=\frac{n_x}{n}\times\Delta \label{eq: true step size}
\end{align}
A Python script is used to prepare the input parameters using the above relations. 

We remark the role of $\Delta$, which represents both length and time in dimensional analysis (recall $c=1$). Thus, we have a degree of freedom (dof) to choose its value without affecting the physics, as if we were changing the length unit. We emphasize this notion because 
there is in fact a deeper reason for having this dof: Eq.~\eqref{eq:double-excitation delay differential eq} is a wave equation, and thus \emph{scale invariant} \cite{PhotonicCrystalBook}. It is well-known that FDTD is quite suitable for scale-invariant problems, and the step size $\Delta$ is just a conceptual quantity for proper discretization and is irrelevant of the physics. 
What matters is the dimensionless quantities such as $(\omega_0\Delta)$ and $(\Gamma\Delta)$, not $\Delta$ itself. 

The current implementation is not yet strictly scale invariant --- we still explicitly keep $\Delta$ as a mandatory parameter, whose value does not affect the result (providing rounding errors are negligible), but it is not necessary if the code is implemented in a different way. What really controls the (relative) step size, and therefore the error, is the ratio of $n_x/n$; see Eq.~\eqref{eq: true step size}. As a result, we need $n_x\gg n$ to have many steps per wavelength. Empirically we find $\Delta/\lambda_0\leq1/100$ is desirable.

\section{Evaluating incomplete Gamma functions on the complex plane}
\label{appen: incom_gamma}
In the present work, the incomplete Gamma function $\gamma(n,z)$ naturally emerges from our equations. While there are several open-sourced math libraries such as GSL and Boost that implement the real-valued version, evaluating $\gamma(n,z)$ for complex-valued argument $z$ is a very tricky task and to our knowledge no open-sourced code provides such a routine. Commercial softwares like Mathematica does provide one for arbitrary arguments, but its efficiency is nonideal for simulation purposes. As a by-product, therefore, we provide an implementation of the incomplete Gamma function for nonzero positive integers $n=1, 2, \cdots$, and complex-valued $z$.

\begin{figure}[htbp]
	\centering
	\includegraphics[trim=2cm 2.2cm 2cm 2.2cm, clip=true, scale=0.7]{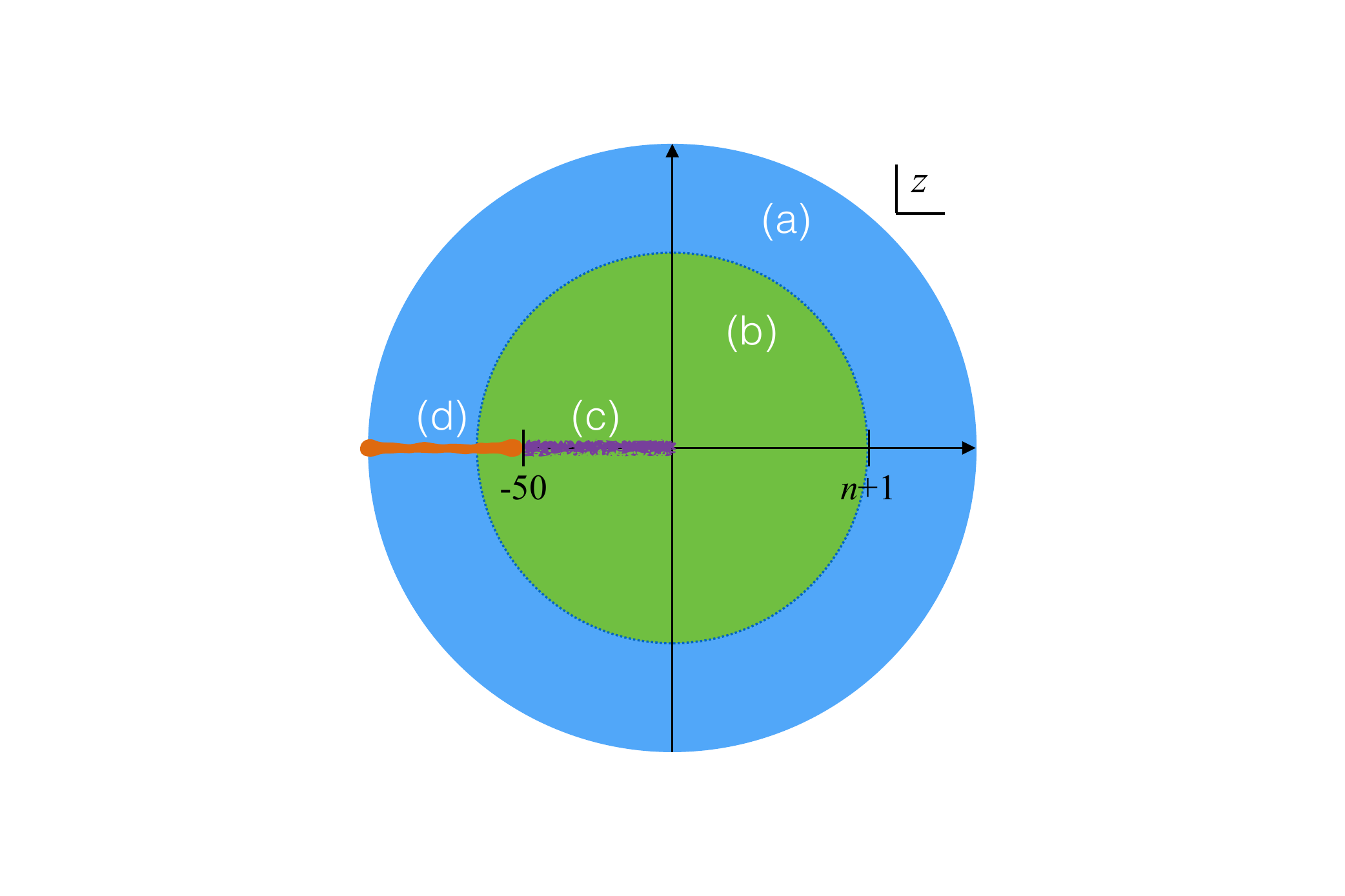}
	\caption{Evaluating $P(n,z)$ for complex-valued $z$. Depending on $|z|$, we use different formulae, labeled from (a) to (d), to achieve fast and accurate convergence. Note that (c) and (d) are employed when $\re(z)<0$ and $|\im(z)|\lesssim10^{-16}$.}
	\label{fig:incomplete Gamma}
\end{figure}

Specifically, we implement the normalized lower incomplete Gamma function \cite{NISThandbook}
\begin{equation}
P(n, z) = \frac{\gamma(n, z)}{\Gamma(n)}=\frac{\gamma(n,z)}{(n-1)!}.
\end{equation}
Other members in the incomplete-Gamma family can be easily obtained if $P(n,z)$ is known. 
The most challenging part is when $z$ is on the negative real axis. Fortunately, this problem is recently tackled in Ref.~\cite{GilACM17}, and we incorporate their findings into our implementation, which works well even when $z$ has a tiny, nonzero imaginary part. Our implementation is summarized in Fig.~\ref{fig:incomplete Gamma}. We evaluate $P(n,z)$ according to four different expressions for $z$ in different regions on the complex plane:
\begin{enumerate}
	\item[(a)] Continuous fraction; see Eq.~(6.2.7) in Ref.~\cite{NumericalRecipes}:
	\begin{equation}
	Q(n, z) = 1-P(n, z) = \frac{e^{-z}z^n}{\Gamma(n)}\left(\frac{1}{z+1-n-}\,\frac{1\cdot(1-n)}{z+3-n-}\,
	\frac{2\cdot(2-n)}{z+5-n-}\cdots\right);
	\end{equation}
	\item[(b)] Series expansion; see Eq.~(6.2.5) in Ref.~\cite{NumericalRecipes}:
	\begin{equation}
	P(n, z) = e^{-z} z^n \sum_{i=0}^\infty \frac{z^i}{\Gamma(n+i+1)};
	\end{equation}
	\item[(c)] Series expansion for $\gamma^*$; see Eq.~(6) in Ref.~\cite{GilACM17}:
	\begin{equation}
	P(n, z) = z^n \gamma^*(n, z) = \frac{z^n}{\Gamma(n)} \sum_{i=0}^\infty \frac{(-z)^i}{i!(i+n)}.
	\end{equation}
	\item[(d)] Poincar\'{e}-type expansion; see Eq.~(29) in Ref.~\cite{GilACM17}:
	\begin{equation}
	P(n, -z) = (-z)^n \gamma^*(n, -z)\sim \frac{(-1)^n e^z}{\Gamma(n)} \sum_{i=0}^\infty (1-n)_i z^{n-i-1};
	\end{equation}

\end{enumerate}
We note that combining (a) and (b) gives the standard approach to real-valued $z$ \cite{NumericalRecipes}, and that for very large $|z|$ the computation may not converge \cite{GilACM17}, but the convergence range is large enough for our purposes.

\section{Non-Markovian measures}
\label{append: NM}
In this Appendix we illustrate the computation of the non-Markovian (NM) geometric measure \cite{LorenzoPRA13} quantifying the single-photon scattering process in this system. Specifically, we consider initially the waveguide has a single-photon exponential wavepacket of the form Eq.~\eqref{eq:single-photon exponential wavepacket}, and calculate two functions, $\mu(t)$ and $\lambda(t)$,\textsuperscript{\ref{note5}} for constructing the geometric measure. The detailed discussion is reported elsewhere \cite{FangNJP18}, and here we simply quote the results. The computation can be performed by setting an arbitrarily positive \texttt{alpha}, \texttt{init\_cond=2}, and \texttt{measure\_NM=1} in the input file.

Denoting $\phi(x, t)$ as the photon wavefunction in the one-excitation sector, 
we can define a complex function $\mu(t)$ as
\begin{equation}
\mu(t) \equiv \int_{-\infty}^{\infty}dx\, \phi^*(x, t) \psi(x, t). 
\label{eq:mu}
\end{equation}
Similarly, we define a real function $\lambda(t)$ as
\begin{equation}
\lambda(t) = \int_{-\infty}^{\infty}dx\, |\psi(x, t)|^2 - |e_0(t)|^2, 
\label{eq:lambda}
\end{equation}
where $e_0(t)$ is given by Eq.~\eqref{eq:exact solution single-photon exponential}.
After calculating $\psi(x,t)$, the program will compute $\mu(t)$, $\lambda(t)$, $e_0(t)$ and $e_1(t)$ [Eq.~\eqref{eq:spontaneous emission solution}], and output the results as plain text. 
The geometric measure is defined as \cite{LorenzoPRA13}
\begin{equation}
\mathcal{N}_\text{geo}=\int\limits_{\frac{d|\det M_t|}{dt}>0} \frac{d|\det M_t|}{dt}\,dt,
\label{eq: geo measure}
\end{equation}
and elsewhere \cite{FangNJP18} we show that for our system $\det M_t = |\mu(t)|^2 \lambda(t)$. 
Therefore, once the two functions $\lambda(t)$ and $\mu(t)$ are known, the geometric measure can be calculated easily.\footnote{$\lambda(t)$ and $\mu(t)$ can also be used to construct other NM measures; detail in progress will be reported elsewhere.}

\begin{figure}[!hbtp]
	\centering
	\includegraphics[scale=0.5]{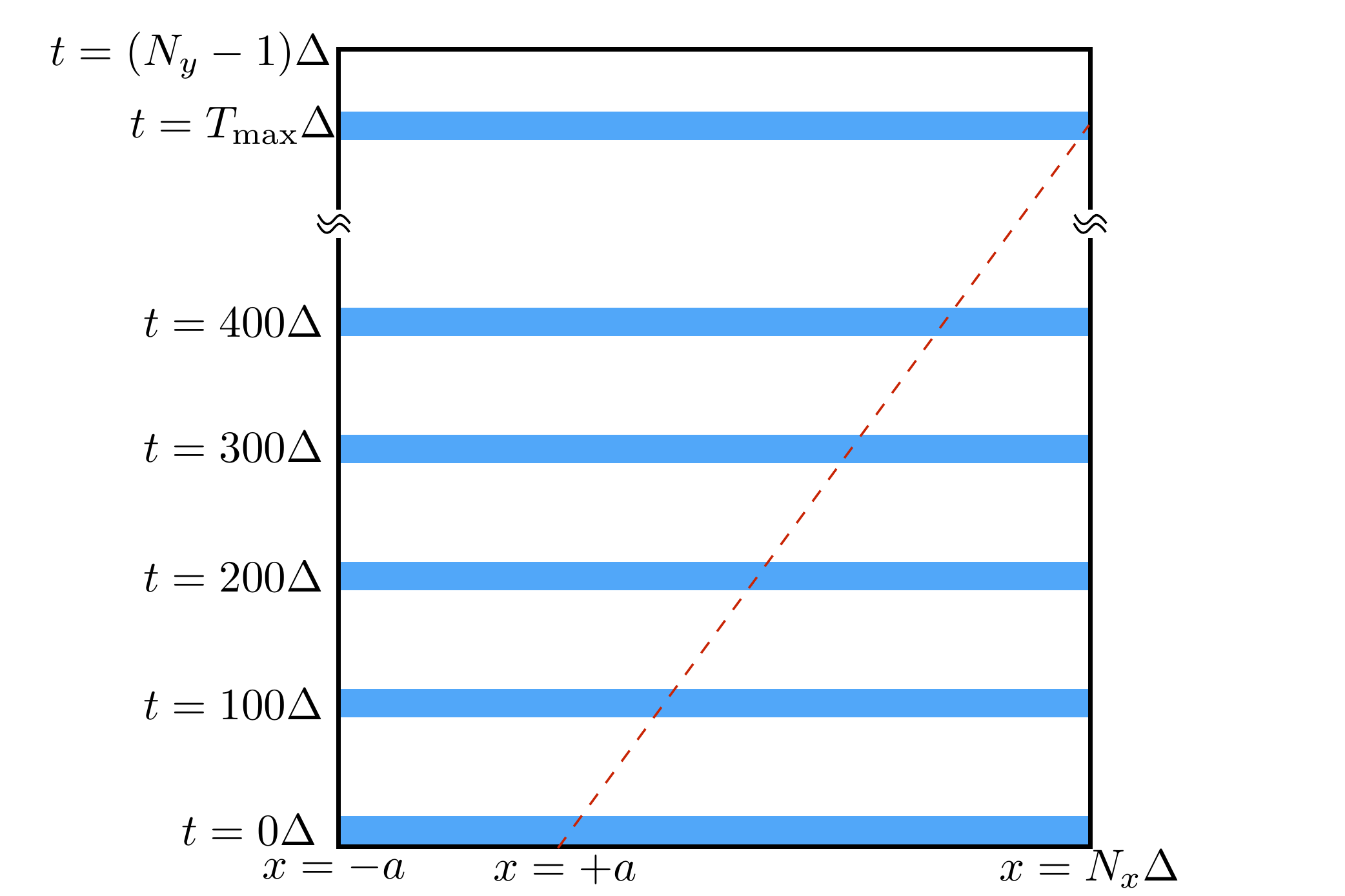}
	\caption{Spacetime layout of the FDTD output files. Information for $x<-a$ is not written to the disk in order to reduce the file size. During post processing, one can load data for every 100 steps into memory (blue stripes). The red dashed line represents the light cone extended from the second 2LS at $x=+a$, and the time at which it intersects with the box boundary is denoted $t=T_\text{max}\Delta$.}
	\label{fig: file processing}
\end{figure}

For users interested in computing the two functions themselves, we note that the spatial integrals in Eqs.~\eqref{eq:mu} and \eqref{eq:lambda} need some care, which is explained in the rest of this section. 
(In our code they are computed according to the trapezoidal rule.)
One of the issues for numerically computing these integrals is that the amount of data generated by the FDTD program can be huge, as previously explained. 
It is certainly inconvenient to write data into a file and subsequently load them into the memory for post processing.

We suggest a manageable way to reduce the disk and memory usages. First, note that the analytical solution in $x<-a$ is known, so do not write data in this region into files. The resulting data layout is shown in Fig.~\ref{fig: file processing}. Doing this will cut half of the disk usage (compare Fig.~\ref{fig: file processing} with Fig.~\ref{fig:FDTD_schematic}). Next, depending on the system time scale, one can selectively load data for every, say, 100 steps (blue stripes in Fig.~\ref{fig: file processing}; set \texttt{Tstep=99}). This will significantly reduce the memory usage. One can keep loading data into the memory until reaching $t=(N_y-1)\Delta$, the max simulation time in FDTD, but for the purpose of calculating the integrals, one may only use data up to $t=T_\text{max}\Delta$, where
\begin{equation}
T_\text{max}=\min(N_y-1, N_x-n_x/2).
\end{equation}
The reason is that the time $T_\text{max}\Delta$ is where the second light cone intersects with the box boundary. If data beyond this limit is used, the wavefront will go outside of the box and the integrals would be underestimated. Thus, instead of positive infinity, numerically the upper bound for those integrals should be the $x$-coordinate that intersects with $t=T_\text{max}\Delta$ at the 2nd light cone.
Finally, we note that \texttt{Tstep} should be wisely chosen so that after interpolation the output data can faithfully represent the result.





\bibliographystyle{model1a-num-names}
\bibliography{WQED_2016,\jobname}

\end{document}